\newif\ifPDF

\documentclass[usenatbib]{mn2e}
\ifPDF
  \usepackage[T1]{fontenc}
  \usepackage{times}
  \usepackage[hypertex,colorlinks=true,citecolor=blue]{hyperref}
  \usepackage{aeguill}
  \usepackage[pdftex]{graphicx,color}
  \usepackage{subfigure}
  \usepackage{afterpage}
\else
  \usepackage[T1]{fontenc}
  \usepackage{times}
  \usepackage[dvips]{graphicx,color}
  \usepackage{subfigure}
  \usepackage{afterpage}
\fi
\title[Results from multi-frequency observations of PSR B0826$-$34]
{Results from multi-frequency observations of PSR B0826$-$34}
\author[B. Bhattacharyya]
{B. Bhattacharyya,$^1$ Y. Gupta$^1$  and J. Gil$^2$\\\\
$^1$National Centre for Radio Astrophysics, TIFR, Pune University Campus, Post Bag 3, Pune 411 007, India\\
$^2$J. Kepler Institute of Astronomy, University of Zielona Gora, Poland }
\date{Accepted. Received}
\begin{document}
\label{firstpage}
\maketitle
\pagerange{\pageref{firstpage}--\pageref{lastpage}} \pubyear{2007}
\def\LaTeX{L\kern-.36em\raise.3ex\hbox{a}\kern-.15em
    T\kern-.1667em\lower.7ex\hbox{E}\kern-.125emX}
\begin{abstract}
We report new results obtained from multi-frequency observations of PSR B0826$-$34 with the 
Giant Metrewave Radio Telescope (GMRT). 
(1) We find no evidence of weak emission during the typical long null state of this pulsar, 
simultaneously at 303 and 610 MHz, as well as individually at 157, 325, 610 and 1060 MHz at 
separate epochs. Our limit of non-detection is at $\sim$ 1\% or better of the peak of the 
active state profile, and corresponds to $\sim$ 2 mJy at 610 MHz. (2) Significant correlation 
in the total intensity of the individual pulses between 303 and 610 MHz is reported from the 
simultaneous dual frequency observations, which is indicative of the broadband nature of the 
emission. We also report correlation between total energy in the main pulse and inter-pulse 
region from the high sensitivity single frequency observations at 610 and 1060 MHz. (3) Though 
we find the drift pattern to be very similar in the simultaneous 303 and 610 MHz data, we 
observe that the drift band separation ($P_2$) evolves significantly between these two 
frequencies, and in a manner opposite to the average profile evolution. In addition, we confirm 
the dependence of $P_2$ on pulse longitude at 303 MHz and find indications for the same at 610 MHz. 
We also present results for subpulse width ($\Delta\Phi_s$) at different frequencies, and as well 
as a function of pulse longitude. (4) As a natural out-come of the simultaneous dual frequency 
observations, we obtain an accurate DM value, equal to  52.2(6) $pc/cm^3$, for this pulsar.

\end{abstract}
\section{Introduction}                                                 \label{sec:intro}       
The phenomenon of subpulse drifting (first reported by \cite{Drake_etal}), is manifested 
as an organised subpulse behaviour $-$ subpulses appear at progressively changing longitude in 
the pulse window following some particular path. The path followed by the subpulses is 
specific to the individual pulsar concerned and is known as drift band. It has been recently 
shown that subpulse drifting may be fairly common among pulsars \citep{Weltevrede_etal_06, Weltevrede_etal_07}. 
Wide profile pulsars showing this phenomenon can be especially useful tools to understand the 
distribution of the radio emission regions above the neutron star's polar cap and the physics 
of the pulsar emission mechanism. Since the wide profiles are usually interpreted as emission 
from one magnetic pole of a highly aligned pulsar (i.e. magnetic axis almost parallel to rotation 
axis), our line of sight samples a large part of the polar cap. This leads to enhanced possibility 
of seeing multiple drift bands in such pulsars, which can provide valuable insights into the pulsar 
magnetospheres, as exemplified by some of the new and substantive results for PSR B0826$-$34 
\citep{Gupta_etal,Esamdin_etal}, PSR B0815$+$09 \citep{Qiao_etal} and PSR B0818$-$41 
\citep{Bhattacharyya_etal}.

PSR B0826$-$34 is a pulsar with one of the widest known pulse profiles. The earlier studies of this 
pulsar \citep{Durdin_etal,Biggs_etal,Gupta_etal} have brought out some unique properties $:$ strong 
evolution of the average profile with frequency, apparent nulling for 70\% of time and a remarkable 
subpulse drift property $-$ multiple, curved drift bands with frequent changes and sign reversals of 
drift rate. 

Multi-frequency observations, especially when done simultaneously at the different frequencies, are 
very useful for the study of various aspects of the frequency dependence of pulsar radiation 
\citep{Taylor_etal_75,Bartel_etal_78a,Bartel_etal_81a,Bhat_etal}. Dispersion measure (DM) of pulsars 
can also be measured from single epoch simultaneous dual frequency observations \citep{Ahuja_etal}. 
Simultaneous, multi-frequency observations of PSR B0826$-$34 can be useful on a variety of counts:
(i) Determination of accurate DM, as the values reported and used in the literature span a fairly 
wide range, from 47 to 65 $pc/cm^3$.
(ii) Study of the pulsar in the null state: \cite{Esamdin_etal} claim the presence of weak emission,
at 1374 MHz, during the long duration nulls of this pulsar. They identify these null states as some 
weak mode of emission where the pulse intensity is 2\% of that of the regular, strong mode, with a 
pulse profile that is similar to the strong mode at lower frequencies. Sensitive, multi-frequency 
observations at the lower frequencies should provide useful information about the behaviour in the
null state.
(iii) Study of simultaneous subpulses: Correlated behaviour of subpulses at different frequencies 
\citep{Taylor_etal_75,Bartel_etal_78a,Bartel_etal_81a,karastergiou_etal,kramer_etal} is useful to 
probe the broadband nature of the emission mechanism. For pulsars with drifting subpulses, evolution 
of $P_2$ (longitude separation between adjacent drift bands) with frequency can also be significant 
\citep{Izvekova_etal,Rankin_etal_05,Smits_etal} 
and a comparison of this with the frequency evolution of other profile parameters can be useful. Earlier 
study of drift properties of PSR B0826$-$34 \citep{Biggs_etal,Gupta_etal,Esamdin_etal} 
report different $P_2$ values at various individual frequencies which do not follow the trend predicted 
by profile evolution with frequency. In addition, for PSR B0826$-$34, the intricate nature of the multiple 
drift bands itself deserves a broadband study. 
   
We observed PSR B0826$-$34 using the GMRT, simultaneously at 303 and 610 MHz, and individually at 
157, 325, 610 and 1060 MHz in total intensity mode. In this paper we concentrate on the results 
from the simultaneous dual frequency observations, and supplement these, where necessary, with 
the results from the single frequencies observations. In Sect.\ref{sec:observations} we cover 
the observations, where a new technique for simultaneous dual frequency pulsar observations with 
the GMRT is presented. Different sub-sections of Sect.\ref{sec:results} describe the data analysis 
and results for (1) determination of DM value from simultaneous dual frequency observations, 
(2) study of the pulsar in its null state and (3) subpulse studies. In Sect.\ref{sec:Discussions} 
we discuss the implications of our findings.
\section{Observations}                                                 \label{sec:observations} 
\subsection {Simultaneous dual frequency observations} \label{sec:dual_freq_obs}
The GMRT consists of an array of 30 antennas, each of 45 m diameter, spread over a region of 25 km 
diameter, operating at 6 different wave bands between 150 and 1450 MHz \citep{Swarup_etal}. Though 
designed to function primarily as a aperture synthesis telescope, the GMRT can be used in an array mode 
by adding the signals from individual dishes, either coherently or incoherently. Furthermore, 
sub-sets of the 30 antennas can be configured completely independently in what is referred to as 
the sub-array mode, to effectively provide more than one ``single dish'', thereby enabling pulsars 
to be observed simultaneously at multiple frequencies. The signals from different antennas and 
observing frequency bands are eventually converted to a baseband signal of 16 MHz band width, which 
is then sampled at Nyquist rate. The observing bandwidth of 16 MHz is divided into 256 spectral 
channels in the FX correlator. These dual polarisation signals from all the antennas are then 
brought to the GMRT Array Combiner (GAC), where they are added coherently to get the phased array 
outputs for each polarisation \citep{Gupta_etal_00}. The GAC has the facility of assigning 
different values of gains to each spectral channel of each polarisation of each antenna, in a 
completely independent manner, before the addition.

In the simultaneous multi-frequency sub-array mode of pulsar observations, the baseband signals from 
antennas in different sub-arrays (operating at different radio frequencies) can be added together in 
the same GAC, and the data can be recorded as a single stream of multi-channel data, after integration 
by the desired amount. The dedispersed data streams for the different frequencies are then extracted 
from this single data stream during off-line processing. Such a technique was employed by 
\cite{Ahuja_etal}, who carried out simultaneous dual frequency observations using the incoherent array 
mode of operation of the GMRT. They utilised the fact that, while traversing the interstellar medium, 
different frequency signals suffer different amount of dispersion delay and as a result the pulses 
arrive at different times (and hence at different positions in pulsar rotational phase) at different 
frequencies. Hence the on-pulse data from different frequencies can be recovered from appropriate 
phase windows in the pulse period, during the off-line analysis. This scheme provides naturally 
synchronised data from all the sub-arrays, without the need for separate, synchronised receivers 
for each sub-array. However, as \cite{Ahuja_etal} have pointed out, this technique works well 
only for cases where the dispersion curves for different radio frequencies, when mapped to the 16 
MHz baseband, do not overlap with each other for all phases of the on-pulse signal. This criterion 
becomes difficult to meet for pulsars with large duty cycle pulses and for large DMs. Specifically, 
in the case of wide profile pulsars like PSR B0826$-$34, this scheme will definitely fail due to 
the fact that the pulses at both frequencies will overlap in the band and it will be impossible 
to extract the stream of single pulse separately at different frequencies. For such pulsars, we 
have devised a new scheme to carry out simultaneous multi-frequency observations.  

As mentioned earlier, the GAC has the facility of assigning to each spectral channel, a different 
value of gain $-$ including zero gain. This allows selectable parts of the band to be processed, while 
blocking the rest of the band, for each antenna. The setting can be done independently for each 
antenna. This ``band masking'' technique allows simultaneous multi-frequency observations of all 
types of pulsars, independent of the combination of values of period, pulse width and DM, without 
having the need for separate, synchronised pulsar receiver chain for each sub array. The trick is 
to split the available baseband bandwidth into non-overlapping sub-bands, each of these parts being 
assigned to the antennas of one particular sub array, using appropriate band mask settings for each 
antenna. For the antennas of a given sub array, we apply nominal gain value for the typical GMRT 
pulsar observations in the chosen part of the band and the rest of the band is masked by applying 
a zero gain value. During simultaneous dual frequency observations, the available band is 
divided into two parts, each part being alloted to one of the observing frequencies. 

Using this technique, we observed PSR B0826$-$34 simultaneously at 303 and 610 MHz with the GMRT, 
on October 26, 2003. We used two sub arrays, one with 5 antennas (sub array\#1), operating at 303 
MHz, and the other with 10 antennas (sub array\#2), operating at 610 MHz. We put more antennas at 
610 MHz to get desired sensitivity, because of the fact that the pulsar is weaker at 610 than 
at 303 MHz. The signals coming from the antennas of each sub array were added coherently in the 
GAC. The total bandwidth used for the observations was 16 MHz, which was divided equally between 
the two frequency bands by masking 8$-$16 MHz of the band for the antennas in sub array\#1 and 
0$-$8 MHz of the band for the antennas in sub array\#2. 
Before observing the pulsar, both the sub-arrays were individually ``phased-up'' by estimating 
the phases with respect to a reference antenna in each sub-array (from the correlator visibilities 
recorded on a point source calibrator) and applying the corrections for these back into the 
correlator hardware. The voltage outputs from the GAC were passed to the phased array pulsar 
receiver where the data were converted to intensity values and intensities from the two polarisations 
added to obtain the total intensity for each of 256 spectral channels, containing information from 
the two sub-arrays. These data were further integrated in time and finally recorded to disk at a 
sampling rate of 1.024 ms. During the off-line analysis we further integrated these data to a final 
time resolution of 4.096 ms, as this was adequate to study this wide profile pulsar. The duration 
of the pulsar observation was about 1 hr, in which we recorded a data stretch of 2123 pulses from 
PSR B0826$-$34, simultaneously at 303 and 610 MHz. In the first 750 of these pulses, the pulsar 
was in the active state and for the remaining time it was in one of its long duration null states. 
\subsection {Single frequency observations}          \label{sec:single_freq_obs}
We observed PSR B0826$-$34 using a large number of GMRT antennas in phased array mode at 157, 325,
610 and 1060 MHz individually at separate epochs. These observations have significantly higher
sensitivity (by a factor of 2 or more) compared to the corresponding observations from the dual
frequency effort, and were carried out primarily to supplement and improve the conclusions obtained
from the simultaneous dual frequency data set. In addition, some of these observations were accompanied
with flux calibration observations using known point sources, which allows for absolute calibration
of the pulsar's flux. A summary of the main parameters of each observing session is given in 
Table. \ref{table:summary}.
\begin{table*}
\begin{minipage}{180mm}
\caption{ Table containing, (a) summary of all the observations presented in this paper, (b) rms values for pulsar 
active and null state, (c) the flux of the peak of the pulsar active state and the non-detection limit for null 
state and (d) the mean flux for the FP, MP and IP regions of pulsar active state for all the observing sessions 
(see Fig. \ref{fig:avp_303_610} for the positions of the MP, OP and IP windows).}
\label{table:summary}
\begin{tabular}{l|c|c|c|c|c|c|c|c|c|c|c|c|c}
\hline
Frequency &State         &No of pulses &No of     &\multicolumn{2}{|c|}{rms$^\bullet$}  &Flux of peak  &Non-detection   &\multicolumn{3}{|c|}{Mean Flux}\\
(MHz)     &(Active/Null) &analysed     &antennas  &\multicolumn{2}{|c|} {}      &              &limit$^\diamond$&\multicolumn{3}{|c|} {}       \\
          &              &             &used      &\multicolumn{2}{|c|}{(mJy)}&  (mJy)       &   (mJy)        &  \multicolumn{3}{|c|}{(mJy)}                  \\
          &              &             &          &   MP   &   OP                     &              &                & FP          & MP$^\odot$  &  IP$^\odot$ \\\hline
303$^\dagger$  & Active  & 750         &    5     &  371   &   28                     & 1687$^\ast$  &     -          &  443$^\ast$ & 754 $^\ast$& 99$^\ast$ \\
               &         &             &          &        &                          &              &                &             &            &             \\
               & Null    & 750         &    5     &   28   &   28                     &     -        &  31$^\ast$     &             &            &            \\
               &         &             &          &        &                          &              &                &             &            &            \\\hline
610$^\dagger$  & Active  & 750         &    10    &   57   &    5                     & 250$^\ast$   &     -          &  119$^\ast$ & 178$^\ast$ & 123$^\ast$  \\
               &         &             &          &        &                          &              &                &             &            &             \\
               & Null    & 750         &    10    &   4    &    4                     &      -       &   5$^\ast$     &             &            &            \\
               &         &             &          &        &                          &              &                &             &            &             \\\hline
157$^\ddagger$ & Active  & 970         &    16    &  427   &   13                     & 1780$^\ast$  &     -          &  370$^\ast$ &  656$^\ast$&  42$^\ast$ \\
               &         &             &          &        &                          &              &                &             &            &             \\
               & Null    & 970         &    16    &  13    &   13                     &      -       &   23$^\ast$    &             &            &             \\
               &         &             &          &        &                          &              &                &             &            &            \\\hline
325$^\S$       & Active  & 940         &    21    &  304   &   5                      & 1330$^\ast$  &      -         &  393$^\ast$ &  763$^\ast$& 62$^\ast$  \\
               &         &             &          &        &                          &              &                &             &            &            \\
               & Null    & 487         &    17    &  5     &   5                      &      -       &   7$^{\ast\circ}$ &          &            &             \\
               &         &             &          &        &                          &              &                &             &            &            \\\hline
610$^\ddagger$ & Active  & 705         &    17    &  85    &   2                      & 410$^{\ast\circ}$ &  -        & 156$^{\ast\circ}$ & 248$^{\ast\circ}$  & 166$^{\ast\circ}$ \\
               &         &             &          &        &                          &              &                &             &            &              \\
               & Null    & 705         &    17    &  2     &   2                      &      -       &    2$^{\ast\circ}$ &         &            &             \\
               &         &             &          &        &                          &              &                &             &            &             \\\hline
1060$^\ddagger$& Active  & 2900        &    21    &  120    &   10                      & 1310$^\ast$  &       -        & 479$^\ast$  & 485$^\ast$ & 989$^\ast$\\
               &         &             &          &        &                          &              &                &             &            &                 \\
               & Null    & 2900        &    21    &   10    &   10                      &      -       &   15$^\ast$    &             &            &             \\
               &         &             &          &        &                          &              &                &             &            &              \\\hline
\end{tabular}

$\dagger$ : Active and null state data recorded from same simultaneous dual frequency observing session \\
$\ddagger$ : Active and null state data recorded from same single frequency observing session \\
$\S$ : Active and null data recorded from single frequency observations from different days \\
$\bullet$ : Value quoted for profile data with time resolution of 4.096 ms \\
$\diamond$ : 3-sigma Non-detection limit for the null state, for profile data with time resolution of 20.048 ms\\
$\ast$ : Flux estimated using standard GMRT system parameters using relative calibration procedure\\
$\circ$ : Flux estimated using absolute calibration with a flux calibrator source\\ 
$\odot$ : Mean flux is calculated up to 50\% intensity lavel for MP and IP\\ 
\end{minipage}
\end{table*}
\begin{figure}   
\begin{center}
\includegraphics[angle=0, width=0.45\textwidth]{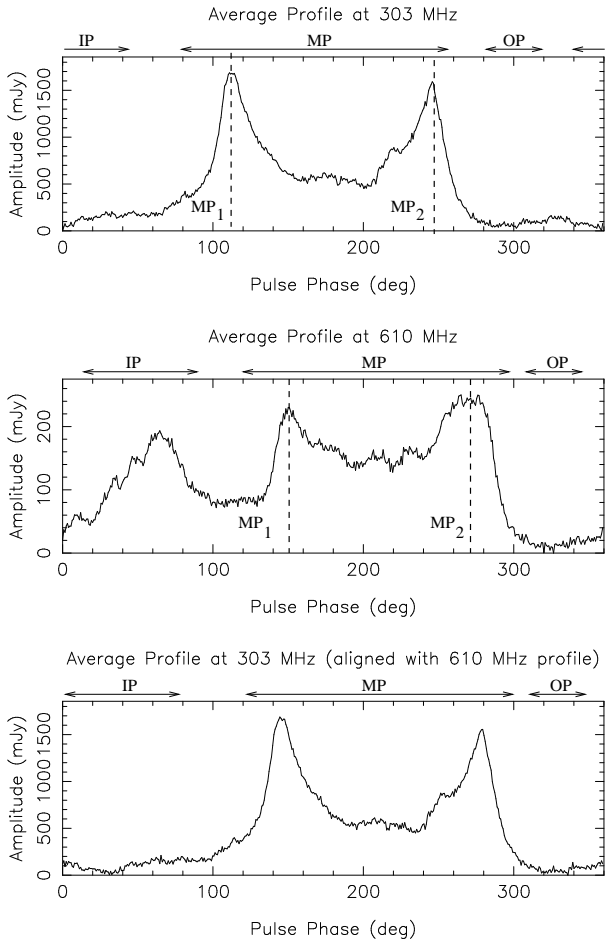}
\caption{Upper two panels plot the average profiles (of the active state) for PSR B0826$-$34 at 303 
and 610 MHz. The third panel shows the 303 MHz profile, aligned to the 610 MHz profile after correction 
for the dispersion delay. MP, IP and OP refer to main pulse, inter-pulse and off pulse, respectively 
(as described in the text); MP$_{1}$ and MP$_{2}$ refer to the two peaks of the MP.}
\label{fig:dm_delay}
\end{center}
\end{figure}   
\section{Data Analysis and Results}      \label{sec:results}   
The simultaneous dual frequency data for PSR B0826$-$34 were dedispersed separately for 303 and 
610 MHz, using only the respective non-masked parts of the baseband signal. The DM value used 
was 52.0, which is close to the final DM value determined by us (see Sect.\ref{sec:DM_determination}). 
The worst kind of radio frequency interference that this data is suffering from, is power line 
related (50 Hz and its harmonics). To remove that, the dedispersed data were put through a radio 
frequency interference filtering routine which detected most (but probably not all) of the power 
line interferences and replaced them by appropriate random noise. The resulting time series 
of first 750 pulses (for which the pulsar was in active state) were then synchronously folded with 
the topocentric pulsar period to generate the average pulse profiles at the two radio 
frequency bands, which are shown in the top two panels of Fig.\ref{fig:dm_delay}. At both frequencies, 
the emission from the pulsar is present over a wide longitude range, as expected. We identify two emission 
regions of the pulse profiles, labeled as main pulse (MP) $-$ roughly from 120\degr to 300\degr pulse 
longitude, and inter-pulse (IP) $-$ roughly form 10\degr to 90\degr, as marked in the panels of 
Fig.\ref{fig:dm_delay}. In addition, we identify an off pulse region (OP) $-$ roughly from 310\degr to 
350\degr pulse longitude. The MP consists of two distinct peaks, denoted by MP$_{1}$ and MP$_{2}$. The 
component separation between these two peaks is 132\degr and 116\degr at 303 and 610 MHz respectively, 
which is in the same ballpark as reported by earlier studies of this pulsar \citep{Biggs_etal,Gupta_etal}.  
\subsection {Determination of accurate DM}                             \label{sec:DM_determination}
In order to compute the dispersion delay between the profiles at the two frequencies, we need to 
establish a fiducial point that is expected to remain intrinsically fixed at a pulse longitude,
even as the pulse profile evolves with frequency. For PSR B0826$-$34, we choose the mid-point of 
the two peaks of the MP of the average profile as the fiducial point. 
The measured time delay $(\Delta t)$ can be expressed as,
\begin{equation} 
\Delta t = (N\times P)+P_{\epsilon}\\
\label{eqn_del_t}
\end{equation}
The topocentric pulsar period $P$ is 1.84881 s. 
$N$ is the integer number of pulsar period delay. Using the observing frequencies, 
the pulsar period and a first guess DM value we estimate $N$ is equal to 1. 
$P_{\epsilon}$ is the delay with in a pulse period. 
$P_{\epsilon}$ is estimated from the position difference of the fiducial point between 303 and
610 MHz pulse profiles. To determine the position of the fiducial points at either of the two
frequencies, we first estimate the position of the two peaks of the MP, by fitting second
order polynomial functions. Here, the errors in the estimation of the location of the
peaks, are smaller at 303 MHz, as compared to the 610 MHz, because of more complex 
structure of the profile peaks at 610 MHz (Fig.\ref{fig:dm_delay}). Using the inferred location 
of the fiducial points at the two frequencies, $P_{\epsilon}$ is estimated as 161.1 ms. The error 
in $P_{\epsilon}$ depends on the error in the estimation of the fiducial point, which follows from 
the error associated with the estimation of the position of MP$_{1}$ and MP$_{2}$ at the individual 
frequencies.\\
Using Eqn. \ref{eqn_del_t} and the well known formula that relates the difference in travel 
time at the two radio frequencies \citep{Backer_etal_93}, we obtain a value of 52.2 $pc/cm^3$ 
for the DM of this pulsar, with an error of $\pm$ 0.6 $pc/cm^3$. The comparison of this result 
with the DM values in the literature is discussed in Sect. \ref{sec:DM_Discussions}.
\subsection {Study of the pulsar in null state}    \label{nullstate}
\begin{figure*}
\begin{center}
\hbox{
  \includegraphics[angle=0, width=0.45\textwidth]{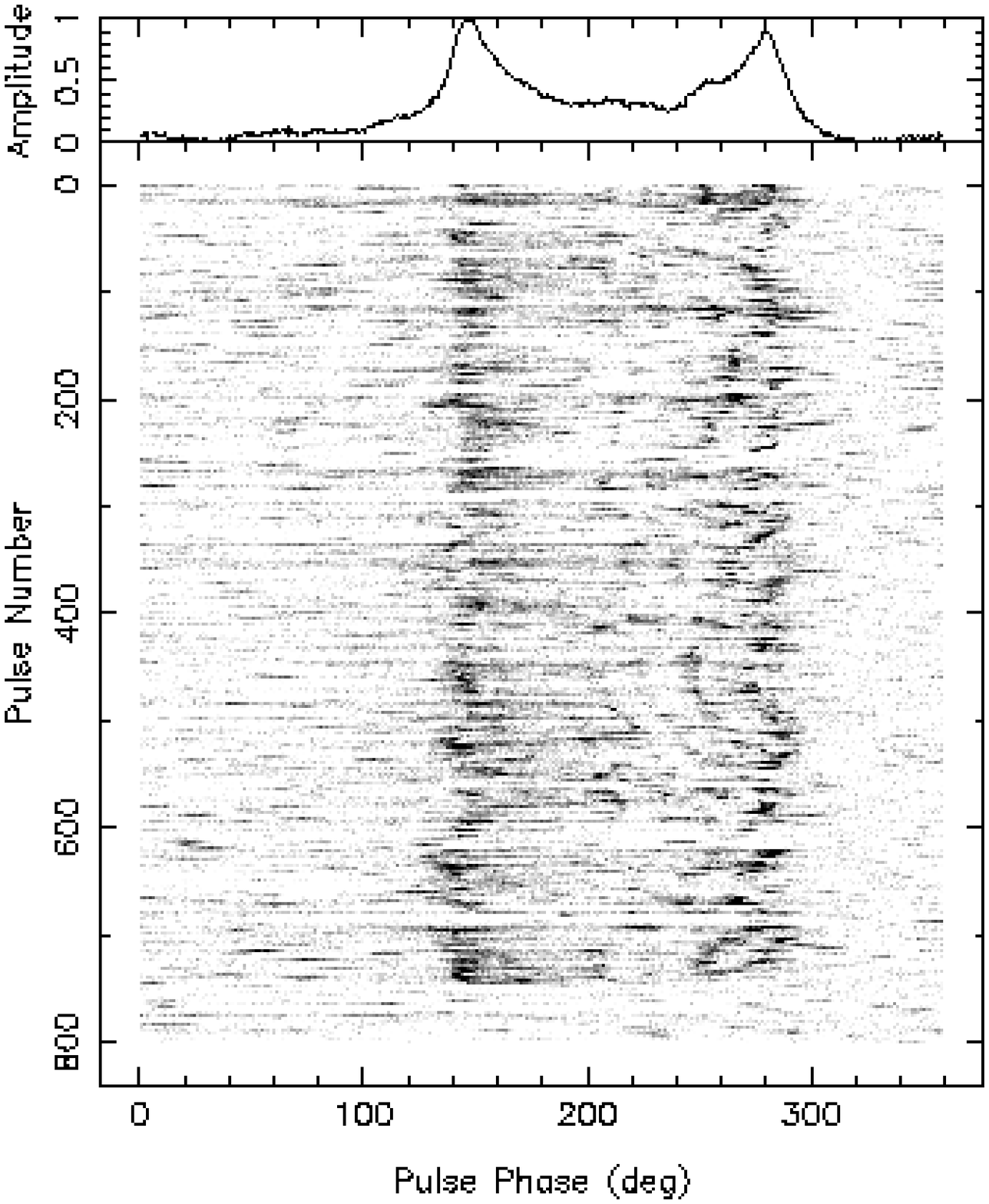}
  \includegraphics[angle=0, width=0.45\textwidth]{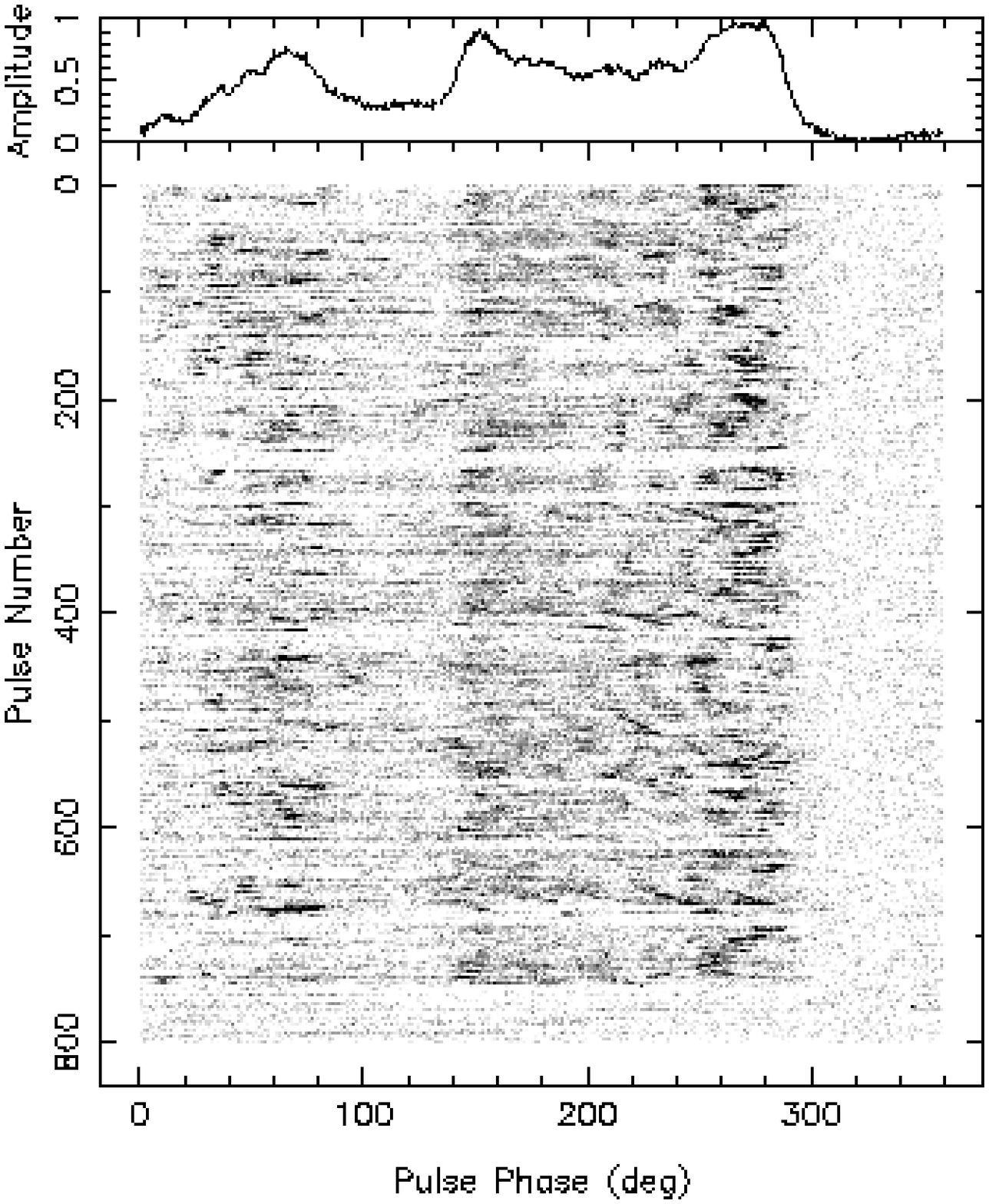}
}
\caption{Gray scale plot of single pulse data for first 800 pulses, from the
simultaneous dual-frequency observations, at 303 (left panel) and 610 MHz (right panel).}
\label{fig:sp_800_303_610}
\end{center}
\end{figure*}

Both short ($\sim$ few pulses) and long duration ($\sim$ few hundred pulses or more) nulls are 
present in our simultaneous dual frequency data. Fig.\ref{fig:sp_800_303_610} shows the subpulse 
patterns at 303 and 610 MHz. For the initial 750 pulses, the pulsar was mostly in an active state 
except for a few short duration nulls which appear to be simultaneous at the two frequencies
(e.g. pulse \# 250 to pulse \# 260, pulse \# 414 to pulse \# 421, pulse \# 613 to 
pulse \# 621 and pulse \# 683 to pulse \# 691, in Fig.\ref{fig:sp_800_303_610}). 
\begin{figure*}
\begin{center}
\hbox{
 \hspace{1cm}
  \includegraphics[angle=0, width=0.4\textwidth]{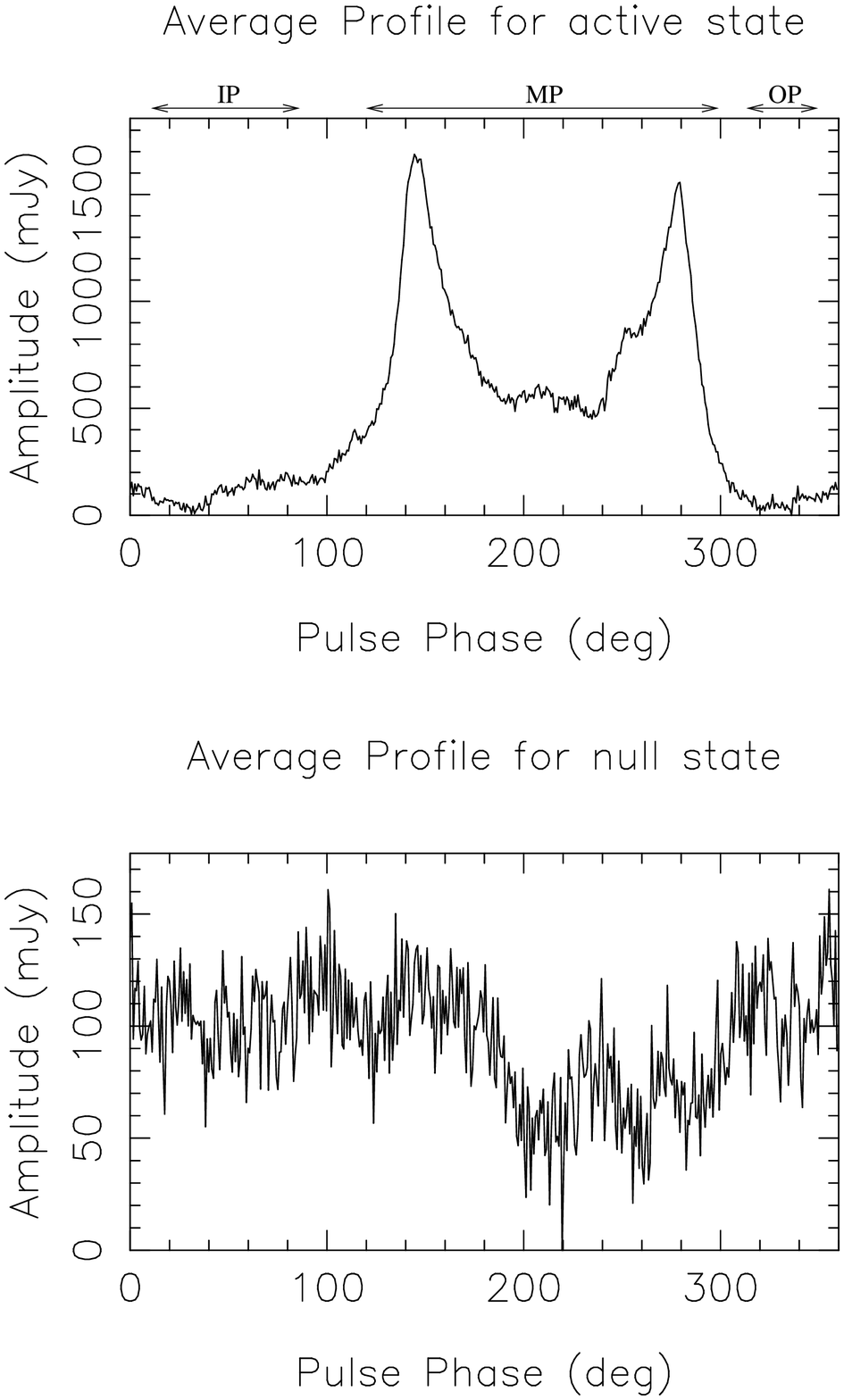}
  \includegraphics[angle=0, width=0.4\textwidth]{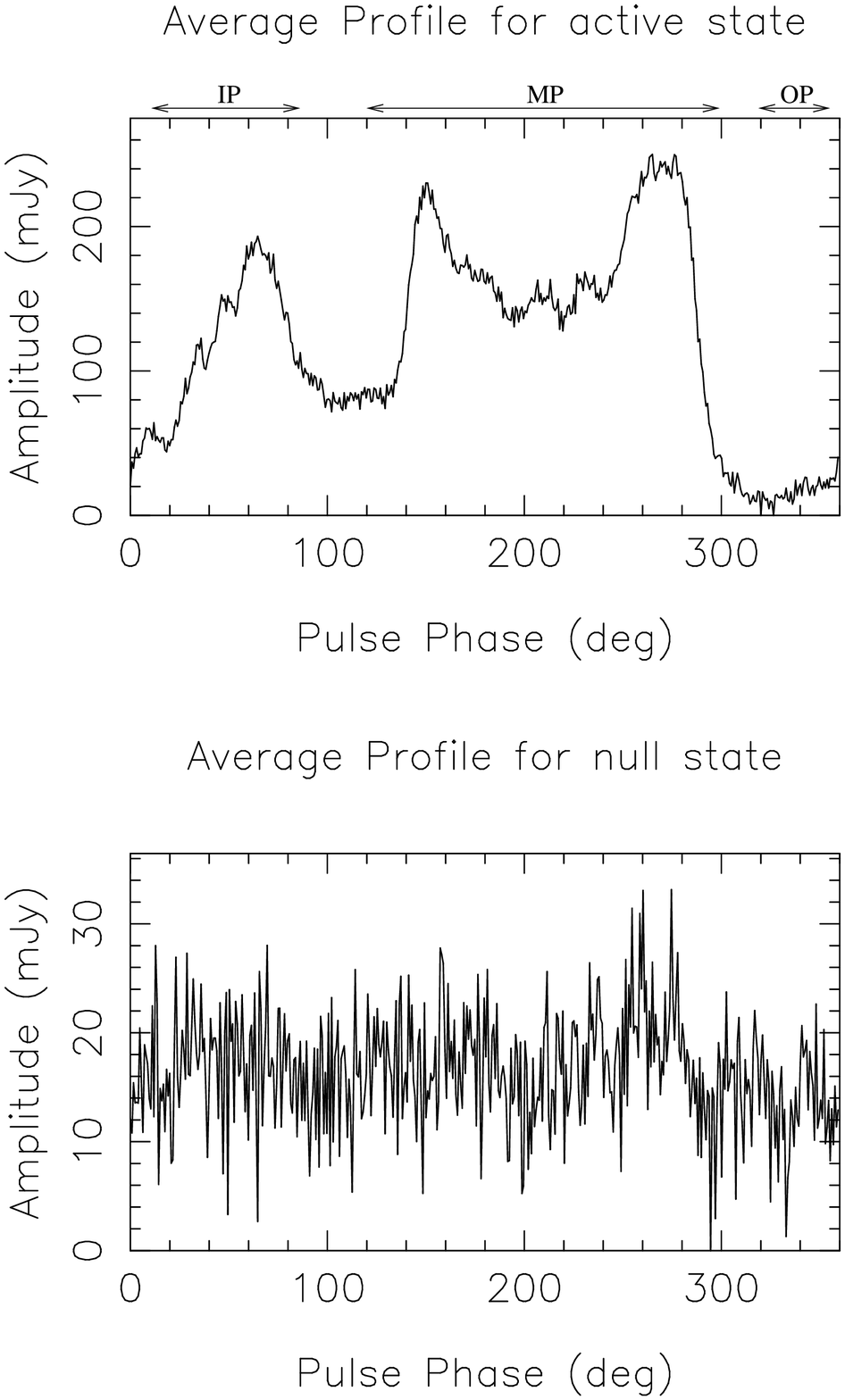}
   }
\caption[Average profile 303 and 610]{Average profiles for PSR B0826$-$34 in the active and null 
states, from the simultaneous dual-frequency observations, at 303 (left panel) and 610 MHz 
(right panel).}
\label{fig:avp_303_610}
\end{center}
\end{figure*}
\begin{figure*}
\begin{center}
\hbox{
 \hspace{1cm}
  \includegraphics[angle=0, width=0.4\textwidth]{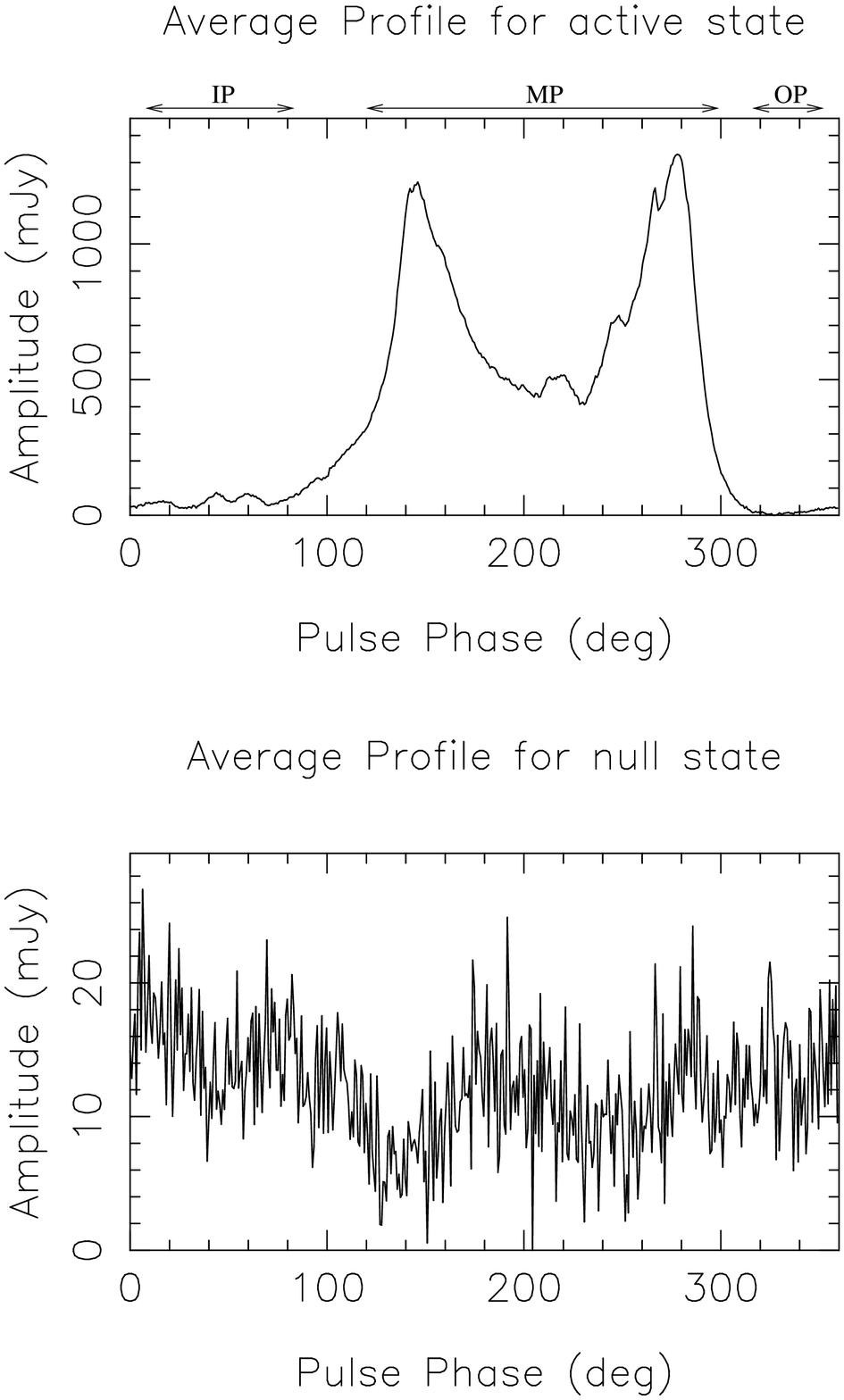}
  \includegraphics[angle=0, width=0.4\textwidth]{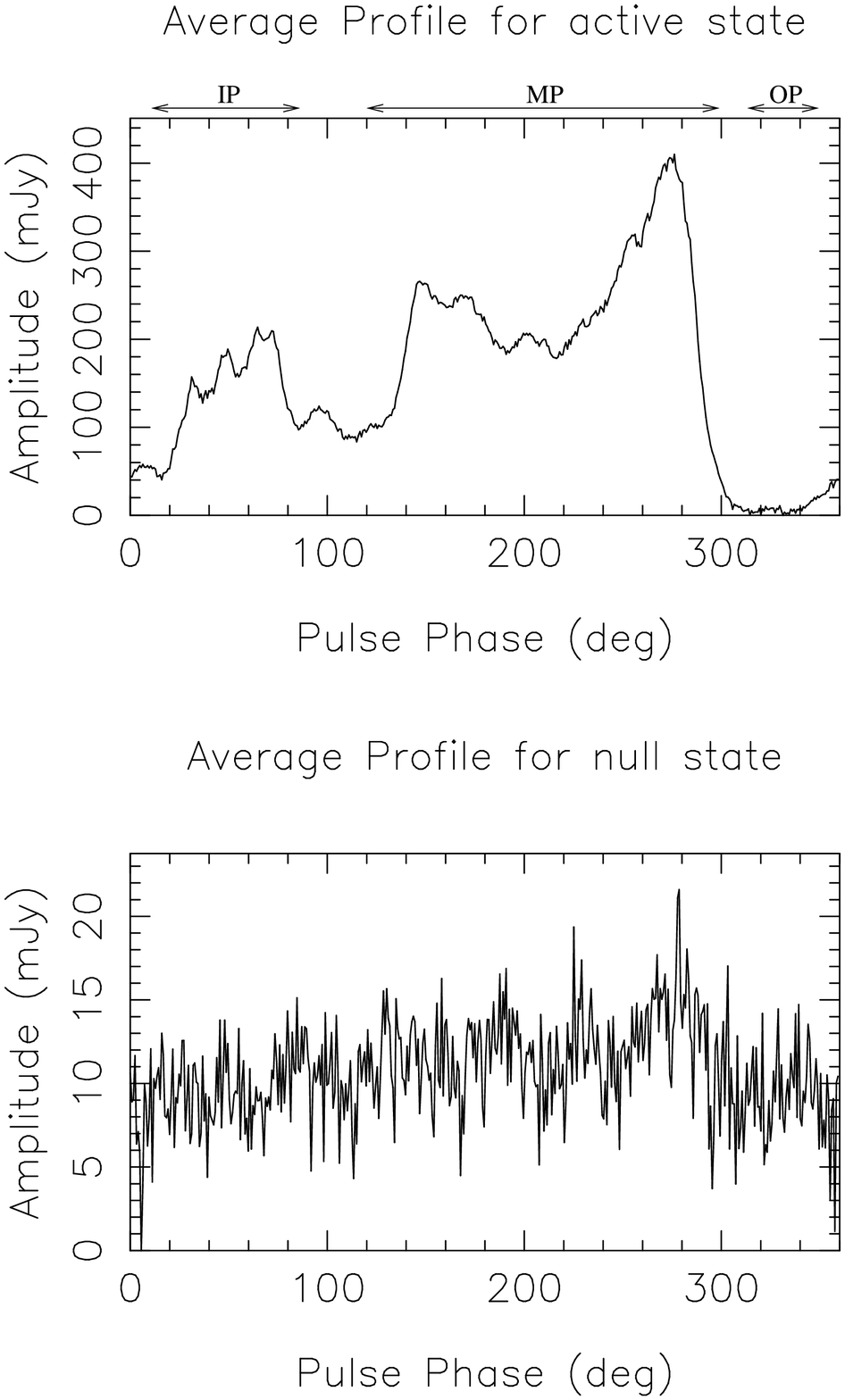}
   }
\caption[Average profile 325 and 610]{Same as Fig. \ref{fig:avp_303_610} for higher sensitivity single frequency observations at 325 (left panel) and 610 MHz (right panel)}
\label{fig:avp_325_610}
\end{center}
\end{figure*}
\begin{figure*}
\begin{center}
\hbox{
 \hspace{1cm}
  \includegraphics[angle=0, width=0.4\textwidth]{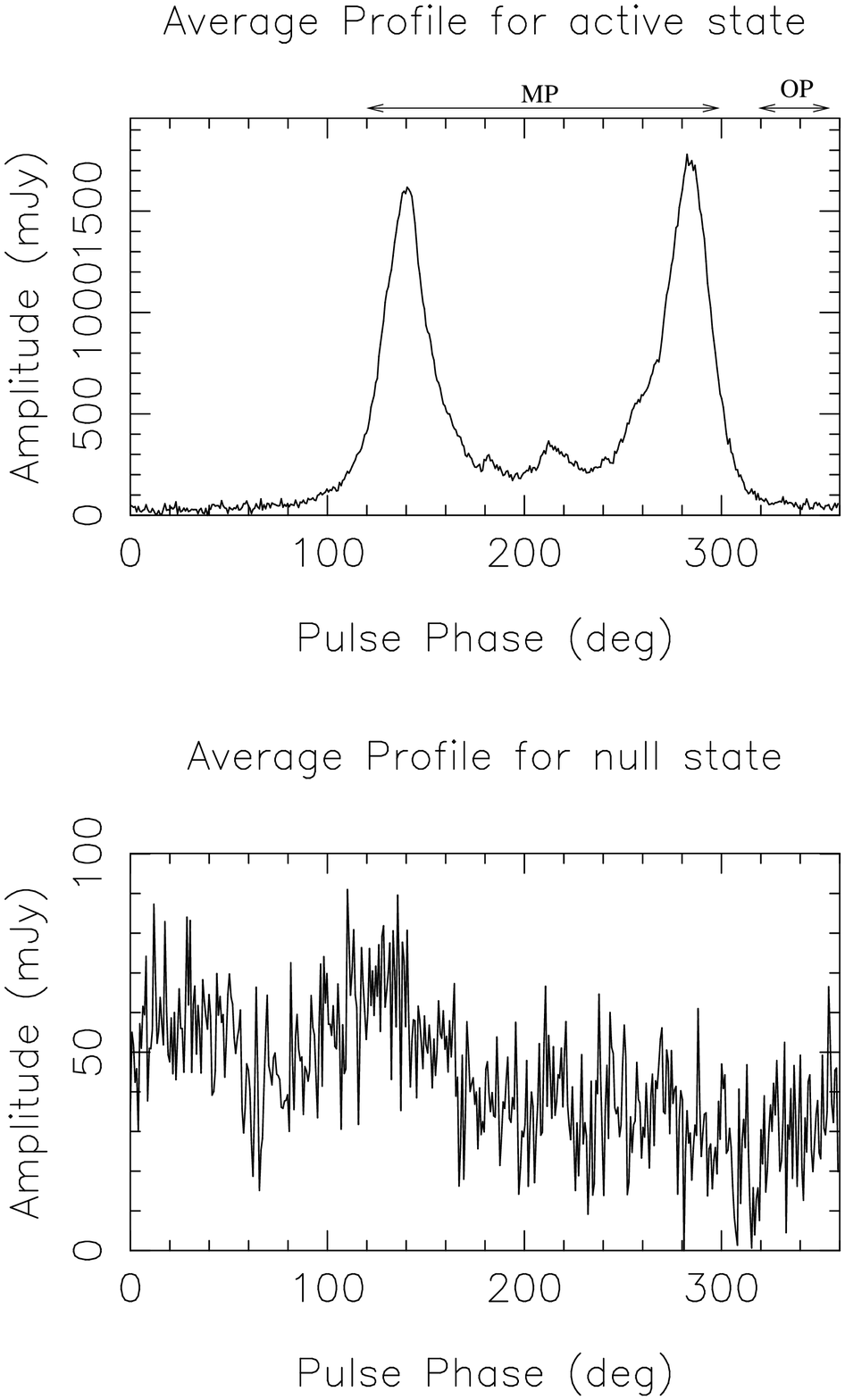}
  \includegraphics[angle=0, width=0.4\textwidth]{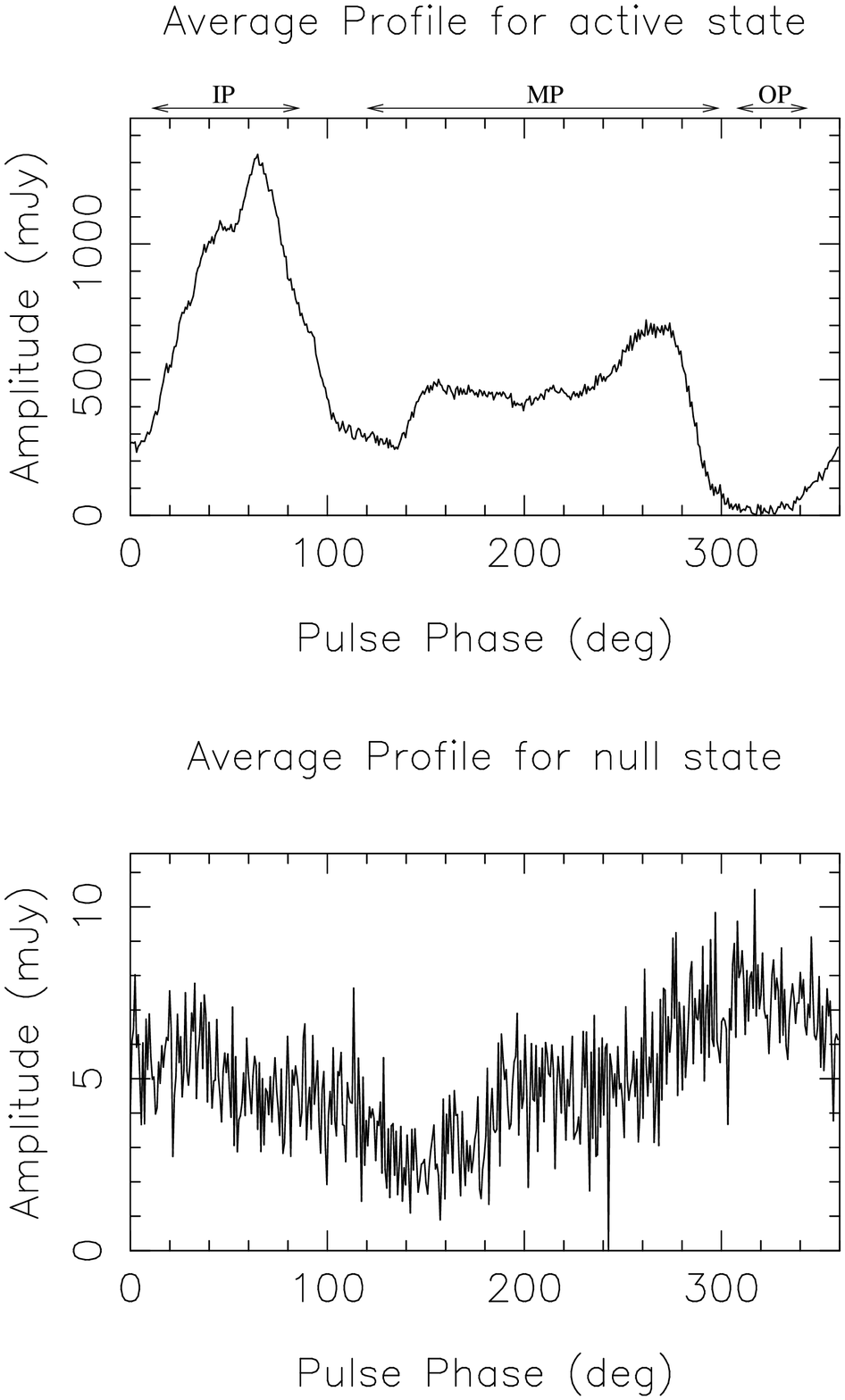}
   }
\caption[Average profile 157 and 1060]{Same as Fig. \ref{fig:avp_303_610} for higher sensitivity single frequency observations at 157 (left panel) and 1060 MHz (right panel)}
\label{fig:avp_157_1060}
\end{center}
\end{figure*}

After the first 750 pulses, the pulsar goes into a long, apparently null state, and this happens 
simultaneously at both the frequencies. The pulsar remains in this state for the rest of the 
observations (about 1373 pulses). We obtain the average profile of the active state by integrating 
the first 750 pulses. Similarly, an average profile of the apparent null state of comparable 
duration is obtained from the next 750 pulses (see Fig.\ref{fig:avp_303_610}). We see that during 
the null state, there is no evidence of any emission that matches with the active state average 
profile for the MP or IP regions, at either of the two frequencies. The mean power level in the null 
state profiles is comparable to the mean level of the OP region of the corresponding active state 
profiles. The fluctuations of the signal level in the null state profiles, though not completely 
Gaussian noise-like (probably due to low-level unexcised interference and other sources of ``red'' 
noise), show no correspondence with the features in the active state profiles. Furthermore, these 
features in the null state profiles have no resemblance or correlation between the simultaneous 
data at 303 and 610 MHz, indicating that these are not genuine profile features. In order to further 
check for any short duration bursts of strong emission in the null state, we divide the data (the 
first $\sim$1500 pulses in the observation) into 20 blocks and compared the average profiles from 
these blocks. For the first 10 blocks, during which the pulsar is in the active state, we see the 
characteristic pulse profile at either of the frequencies. For the next 10 blocks, during which 
the pulsar is in the null state, we do not see any systematic signature of emission, at either 
of the two frequencies, in any one of the 10 blocks. 

For a more thorough investigation of the null state, we compare the rms values for the MP, OP and 
full pulse (FP)  windows of the integrated pulse profiles (see Table. \ref{table:summary}). The rms during 
the pulsar null state is similar for all the windows (MP, OP and FP) and is also similar 
to the rms of the OP region of the pulsar active state. These rms values are much smaller than the rms 
of the MP window of the active state, and their ratios to the corresponding mean values are, to 
first order, consistent with thermal noise statistics. This behaviour is consistent for both the 
frequencies, indicating that the null state signal is similar to the OP region of the pulsar's
active state, and contains no detectable level of emission from the pulsar. 
To further constrain the above conclusion, we integrated the profile data to a time resolution of 
20.048 ms, which is a reasonable compromise between integrating all of the pulsar's signal into
a few bins and retaining enough time resolution to detect any large scale emission structure. 
Comparing the peak deflection in the active state profiles with three times rms of the fluctuations 
in the null state profiles, we conclude that there is no emission in the null state down to a level 
of $\sim$ 2\% of the peak in the active state, for either frequency. 

The dual frequency observations used a limited number of antennas at each frequency and are not
very sensitive. In order to achieve better sensitivity, we use the results from the single 
frequency observations carried out with larger number of antennas in phased array mode.
Fig. \ref{fig:avp_325_610} and Fig. \ref{fig:avp_157_1060} show the active and null state profiles
from these observations. Whereas the active and null state profiles for the 325 MHz observations
are from different days, those for the 157, 610 and 1060 MHz observations are cases where the 
pulsar transited from the active to null state during the same observing session.  
  
As can be seen, there is no evidence for systematic emission from the pulsar in the null state
at any of the frequencies. Non-random features in the null state profiles, where seen, have 
no correspondence between different frequencies, or even for the same frequency at different 
epochs. 
 
Comparing the peak deflection in the active state profile with three times the rms of the 
fluctuations in the null state average profile at a time resolution of 20.048 ms, we conclude 
that at 610 MHz there is no emission in the null state down to a level of $\sim$ 0.5 \% of the 
peak in the active state.  
At 1060 MHz no signature of emission is seen in the null state profile down to $\sim$ 1.2\% of
the peak in the active state. Similarly, at 157 MHz the limit is $\sim$ 1.3\%.

From the absolute flux calibration available, we determine the flux of the peak for 610 MHz 
active state to be $\sim$ 410 mJy, and the 3-sigma non-detection limit in the null state 
corresponds to about 2 mJy. Similarly, from the absolute flux calibration available for 
the null state observations at 325 MHz, we find the 3-sigma non-detection limit at this 
frequency to be $\sim$ 7 mJy. For the cases where we do not have any flux calibrator source 
observed, we follow a relative calibration procedure, using the knowledge of the observing 
parameters and the background sky temperature near the source. As a crosscheck we find that, for 
the single frequency 610 MHz observations, the flux values estimated with absolute and relative 
calibration are similar. 

Though the scintillation properties of PSR B0826$-$34 is not studied, still we try to understand 
the effect of scintillation on the observed flux, by comparing with two nearby pulsars for which the 
scintillation properties are known. The mean flux densities for two nearby pulsars, PSR B0823+26 and 
PSR B0834+06, are observed to change by $\sim$ 35\% between the observing epochs \citep{Bhat_etal}. 
Between the two observing sessions of PSR B0826$-$34 at 610 MHz, we observe, 27\% change of the mean flux 
of the FP. The mean flux at 610 MHz is close to the mean flux estimated by \cite{Biggs_etal} at 645 MHz. 
Estimated mean flux changes by 12\% between two observing sessions at 303 and 325 MHz. So, the flux 
variation observed by us are well within the 3 $\sigma$ limit of what is reported for the nearby 
pulsars. Hence, it is likely that the estimated mean flux values of PSR B0826$-$34 are affected by 
the scintillation to a moderate extent.

\begin{figure}
\begin{center}
\includegraphics[angle=-90, width=0.5\textwidth]{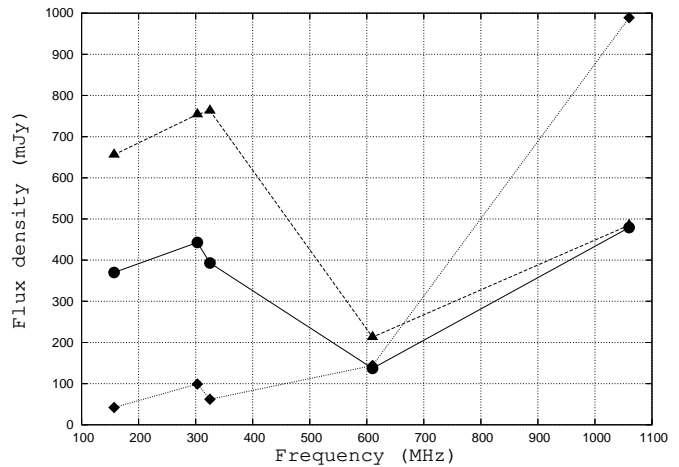}
\caption{Spectra from calculated mean flux densities (Table. \ref{table:summary}) for FP (solid), MP (dashed), IP (dotted)}
\label{fig:flux}
\end{center}
\end{figure}

In Table.\ref{table:summary} we list the flux at the peak and the mean flux of PSR B0826$-$34 for FP, 
MP and IP regions at different frequencies. 
Fig. \ref{fig:flux} illustrates the frequency evolution of the mean flux density of the FP, 
MP and IP regions. At lower frequencies like 157 MHz, the IP is almost absent and the MP is the dominant 
component. With increasing frequency the mean flux density of the IP increases. At 610 MHz, the mean 
flux density of the MP and the IP are almost equal and at 1060 MHz, the IP flux density dominates the MP. 
For the simultaneous dual frequency observations, we see that the FP and MP has quite steep 
spectral index, $\alpha_{fp}$$\sim$ -1.9 and $\alpha_{mp}$$\sim$ -2.1 ,whereas the IP 
has flatter spectral index, $\alpha_{ip}$$\sim$ 0.3 (where $\alpha$ is defined by $I\propto \nu^{\alpha}$). 
These are similar to the results reported by \cite{Biggs_etal} between 408 and 645 MHz. 
From single frequency observations at 610 and 1060 MHz, the calculated spectral indexes are, 
$\alpha_{fp}$$\sim$ 2, $\alpha_{mp}$$\sim$ 1.2 and $\alpha_{ip}$$\sim$ 3.2. Hence, the spectral 
index $\alpha$ for this pulsar is different for different profile components and also evolves with 
frequency. This can be understood by considering that the IP emission is coming from a second ring of emission 
which is missed by our line of sight at the lower frequencies and is visible at the higher frequencies, as 
proposed by \cite {Gupta_etal}.

To summarize, our results show similar limits ($\sim$ 1\% of peak of active state profile, 
or better) of non-detection of emission in the null state, at frequencies of 157, 303, 325, 
610 and 1060 MHz. These results are in apparent disagreement with those obtained at 1374 MHz 
by \cite{Esamdin_etal}, who report presence of weak emission in the apparent null state.  
We discuss the implications of this in Sect.\ref{null_Discussion}.
\subsection{Study of subpulse emission}                                \label{subpulse_study} 
We align the 303 and 610 MHz data from the simultaneous dual frequency observations, by removing 
the initial 2009 ms of data (corresponding to the dispersion delay) from the 303 MHz data. This 
allows us to directly compare the corresponding single pulses from these two frequencies.
\begin{figure}
\begin{center}
\includegraphics[angle=-90, width=0.5\textwidth]{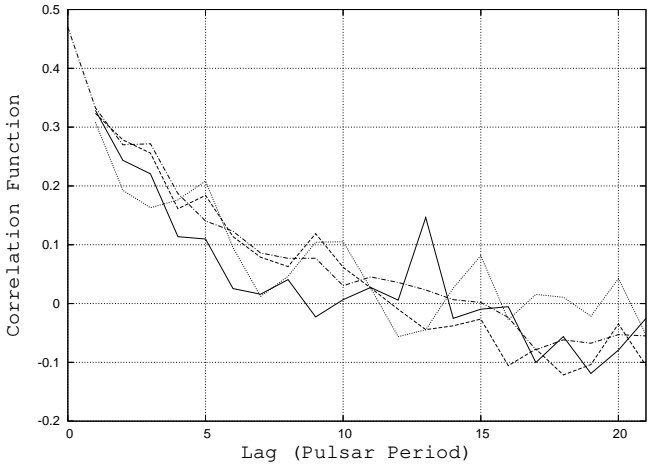}
\caption{Plot of correlation function versus pulse lag: autocorrelation function of TE$_{mp}$ for 303 MHz (solid), same for 610 MHz (dashed), autocorrelation function of TE$_{ip}$ at 610 MHz (dotted), crosscorrelation functions of the TE$_{mp}$ between 303 and 610 MHz (dash-dot). The plot for the autocorrelation function starts from unit lag for all the three cases.}
\label{fig:cor_totenergy_mp}
\end{center}
\end{figure}
\begin{figure*}
\begin{center}
\hbox{
  \includegraphics[angle=0, width=0.45\textwidth]{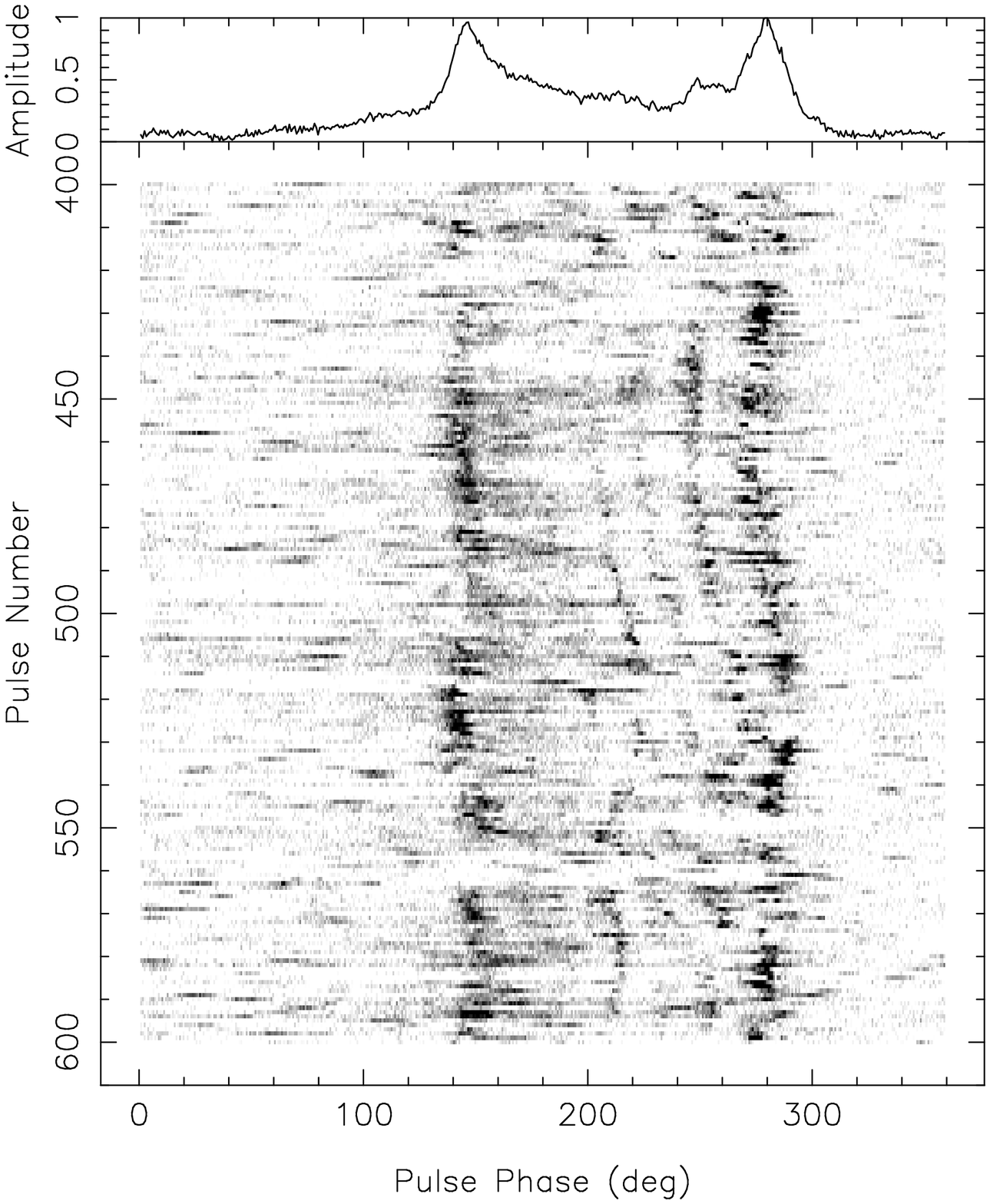}
  \includegraphics[angle=0, width=0.45\textwidth]{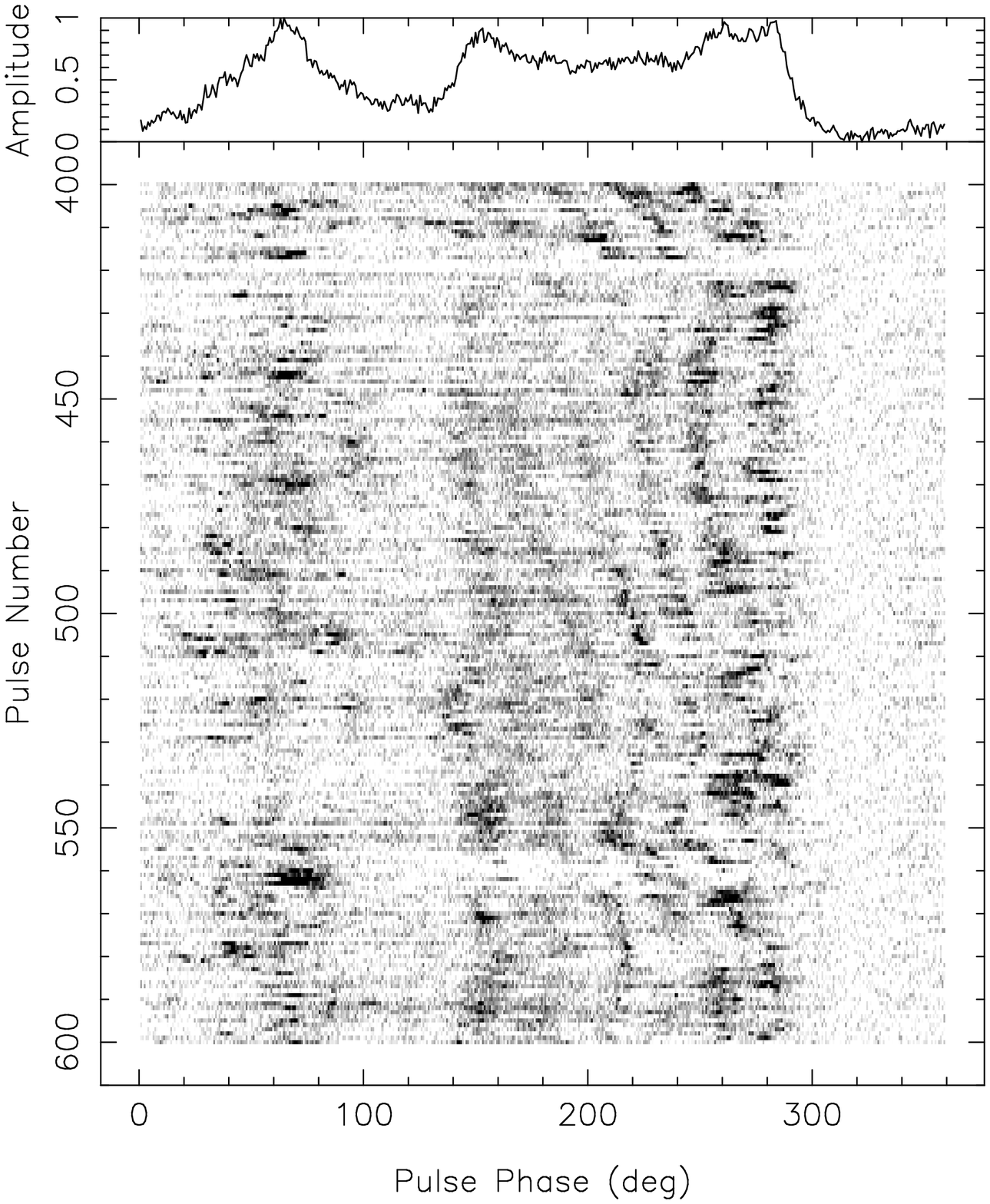}
}
\caption{Gray scale plot of single pulse data for pulse number 400 to 600, from the simultaneous
dual-frequency observations, at 303 (left panel), and 610 MHz (right panel).}
\label{fig:sp_400_600_303_610}
\end{center}
\end{figure*}
\subsubsection{Pulse energy correlation}                               \label{intensity correlation study}
To study the correlation of pulse energy fluctuations as a function of frequency and pulse lag, 
we compute the correlation of the total energy under the MP (hereafter TE$_{mp}$), for different 
pulse lags, individually at 303 and 610 MHz, as well as between these two frequencies. 
For a train of single pulses with pulse number $k$, $1\leq k \leq M$, the correlation function 
is given by \citep{Bartel_etal_81b},
\begin{equation}    
C(l)= \frac{{\frac{1}{M-l}}\sum_{k=1}^{M-l} (I_{f_i}(k)-\langle{I_{f_i}}\rangle) (I_{f_j}(k+l)-\langle{I_{f_j}}\rangle)} {({\sigma_{f_i}^2\sigma_{f_j}^2})^{1/2}}
\label{eqn4}
\end{equation}   
where $i$, $j$ denote the concerned frequencies (303 or 610 MHz); $I_{f_i}(k)$ and $I_{f_j}(k)$ are 
the TE$_{mp}$ of the $k$ th pulse for frequencies $i$ and $j$, respectively; $\langle{I_{f_i}}\rangle$,
$\langle{I_{f_j}}\rangle$ are the mean energies, and $\sigma_{f_i}$, $\sigma_{f_j}$ are the variances 
for TE$_{mp}$ over $M$ pulses. A similar procedure is adopted for the IP, where applicable. 
\begin{figure*}
\begin{center}
\hbox{
  \includegraphics[angle=0, width=0.45\textwidth]{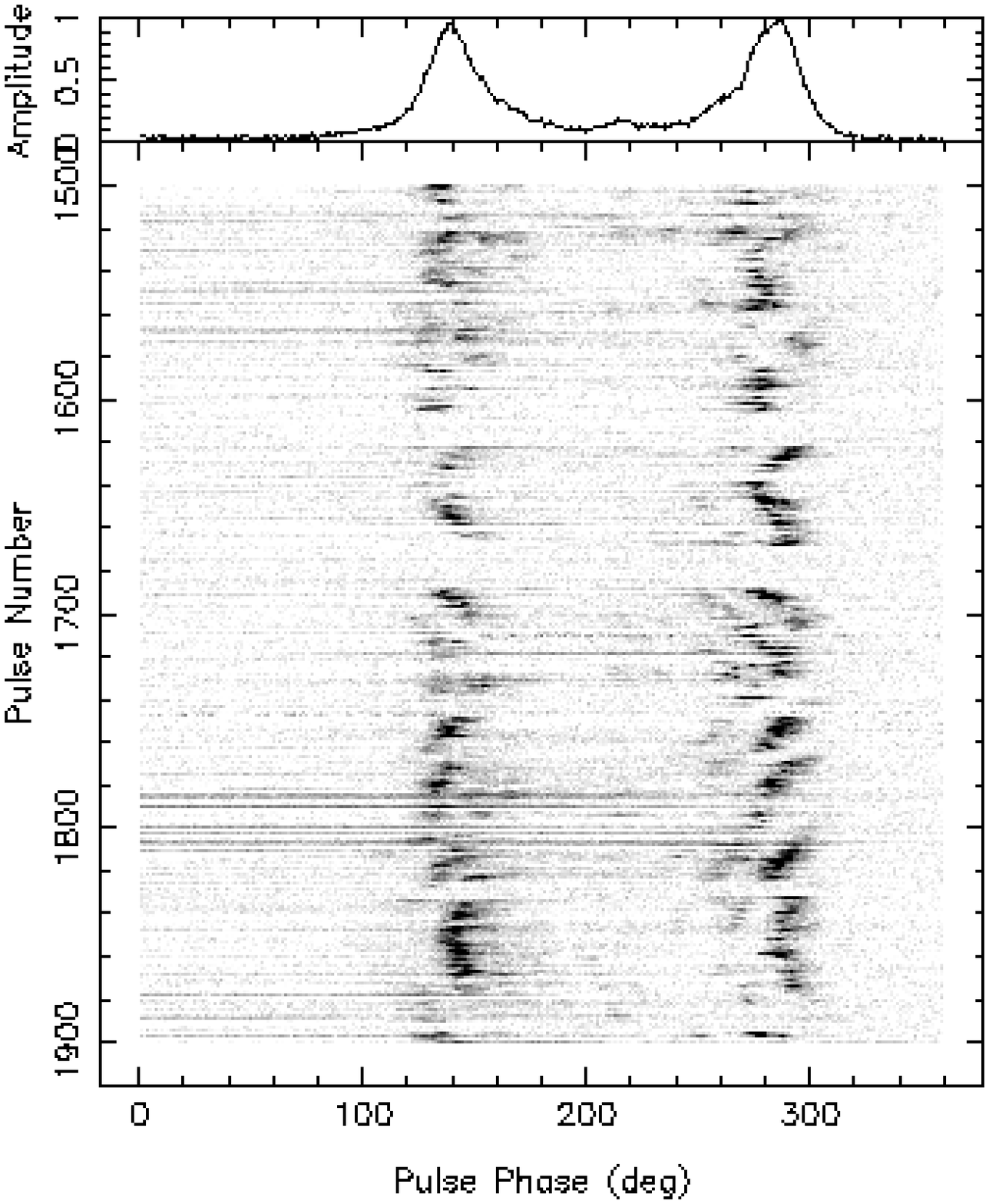}
  \includegraphics[angle=0, width=0.45\textwidth]{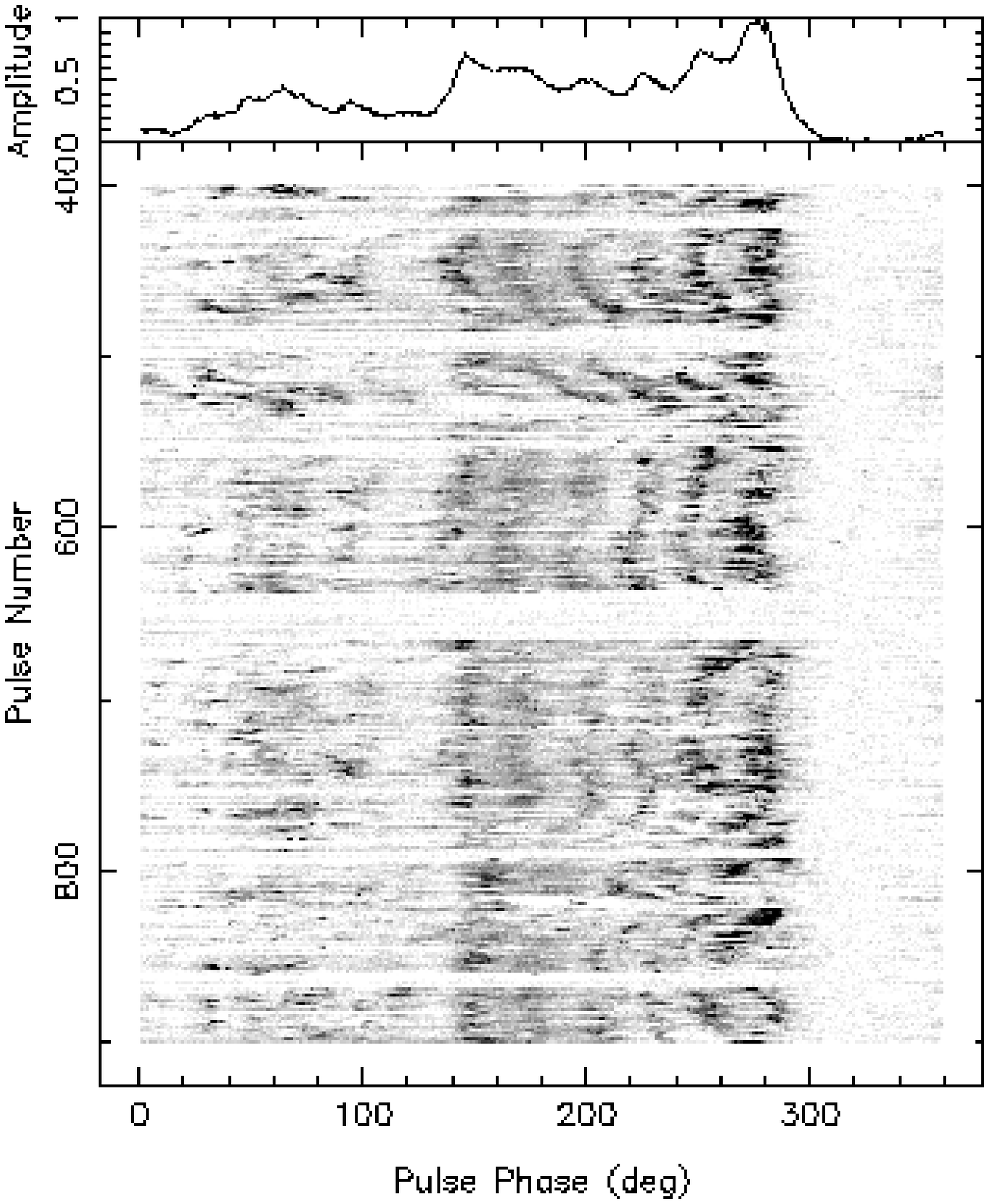}
}
\caption{Gray scale plot of single pulse data from single frequency observations at 157 (left panel) and 610 MHz (right panel).}
\label{fig:sp_full_157_610}
\end{center}
\end{figure*}

Fig.\ref{fig:cor_totenergy_mp} shows the correlation function (for pulse number 1 to 750), for 
autocorrelation of TE$_{mp}$ at 303 and 610 MHz, autocorrelation of TE$_{ip}$ at 610 MHz and 
crosscorrelation of TE$_{mp}$ between 303 and 610 MHz, as a function of increasing pulse lag. 
The TE$_{mp}$ for the active state (pulse number 1 to 750), with zero pulse lag (i.e. $l=0$), is 
47\% correlated between 303 and 610 MHz. Moreover, this level is very similar to the correlation
at each of the two frequencies. All these three curves (as well as that for the autocorrelation of
the IP at 610 MHz), follow very similar shapes with pulse lag, and show positive correlations out
to fairly large lags: about 11 periods. Similar results are obtained from correlation studies 
of the higher sensitivity single frequency data. 
Furthermore, from the single frequency data, we find a crosscorrelation between TE$_{mp}$ and TE$_{ip}$ 
of about 22\% at 610 MHz and 17\% at 1060 MHz. In the autocorrelation function for 303 MHz
(Fig. \ref{fig:cor_totenergy_mp}), and also in the autocorrelation of the 610 MHz data from the
single frequency observations (not shown in Fig. \ref{fig:cor_totenergy_mp}) a secondary maximum is 
present at pulse lag $l=13$, which is not consistently seen in all correlation functions. These 
different aspects of the correlation results, including comparison with existing results for 
other pulsars, are covered in Sect.\ref{subpulse_Discussion}.
\subsubsection{Study of subpulse drifting}                             \label{subpulse drift study}
Drifting subpulses observed simultaneously at 303 and 610 MHz (e.g. Fig.\ref{fig:sp_800_303_610}),
broadly reproduce the features reported by earlier studies at various individual frequencies 
\citep{Biggs_etal,Gupta_etal,Esamdin_etal}. For a closer investigation 
of the simultaneous subpulse behaviour, we zoom into a sequence of 200 pulses (pulse \# 400 to 
pulse \# 600) as shown in Fig.\ref{fig:sp_400_600_303_610}. At any given time, the simultaneous 
multiple subpulses present in the main pulse window follow the same drift rate and sign, at both 
the frequencies. The pulse regions showing different drift rates of opposite signs are always 
connected by a region showing smooth transition of drift rate, as reported by \cite{Gupta_etal}. 
Such transitions, seen in pulse numbers 440 to 460, 500 to 520, 540 to 560 and 570 to 590, also
occur simultaneously at 303 and 610 MHz. The gray scale plot of the single pulses from the higher 
sensitivity 610 MHz single frequency observations (Fig. \ref{fig:sp_full_157_610}) show coherent drifting in 
the MP and IP regions, with approximately 6 drift bands in the MP and 4 drift bands in the IP.  
Subpulse drifting is observed under the two peaks of the MP at 157 MHz (Fig. \ref{fig:sp_full_157_610}). 
\begin{figure}
\begin{center}
\includegraphics[angle=-90, width=0.5\textwidth]{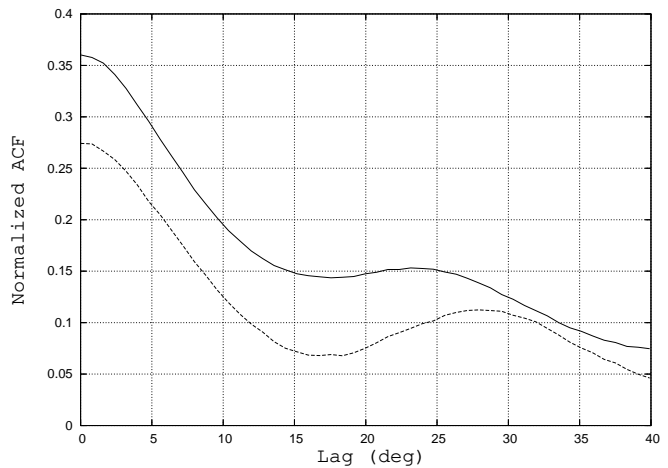}
\caption{Autocorrelation results for section of profile for the MP (longitude 120\degr to
300\degr), calculated with one pulse offset, for pulse \# 1 to 750 at 303 (solid)
and at 610 MHz (dash). Secondary maximum of the autocorrelation function for 303 MHz
is 22.6\degr and 610 MHz is 27.9\degr.}
\label{fig:ccf_303_610}
\end{center}
\end{figure}
Though the drifting subpulses show very similar structure in our simultaneous dual frequency
observations, there are subtle differences such as changes in the separations and widths of 
the subpulse as a function of frequency. To investigate the change of drift band separation, we
compute $P_2$ from the autocorrelation function of the single pulses, averaged over the 750 pulses 
in the active state.  In order to clearly see the secondary maximum in the correlation function
(which will give us the $P_2$ value), we correlate adjacent pulses with different longitude lags, rather 
than the same pulse (which is the traditionally correct procedure). This one pulse offset method
suppresses a large amount of the correlation within the same subpulse, and as a consequence the 
secondary peak becomes stronger and easily detectable. This can be seen by comparing
Fig.4 of \cite{Gupta_etal} with our Fig.\ref{fig:ccf_303_610}. We find $P_2$ = 22.6$\pm$0.8\degr 
at 303 MHz and 27.9$\pm$0.8\degr at 610 MHz. Thus, the value of $P_2$ evolves with frequency. 
Any doubt about epoch dependent variations of $P_2$ manifesting as variation of $P_2$ with 
frequency can clearly be ruled out with these simultaneous dual frequency results.
Further frequency evolution of $P_2$ is obtained from our single frequency observations at 
different frequencies, the results from which are summarised in Table. \ref{table:avp_p2}. 
In addition, we find $P_2$ $\sim$ 20\degr for the IP region, from 610 and 1060 MHz single frequency
observations. 
 \begin{figure*}
\begin{center}
\hbox{
  \includegraphics[angle=-90, width=0.45\textwidth]{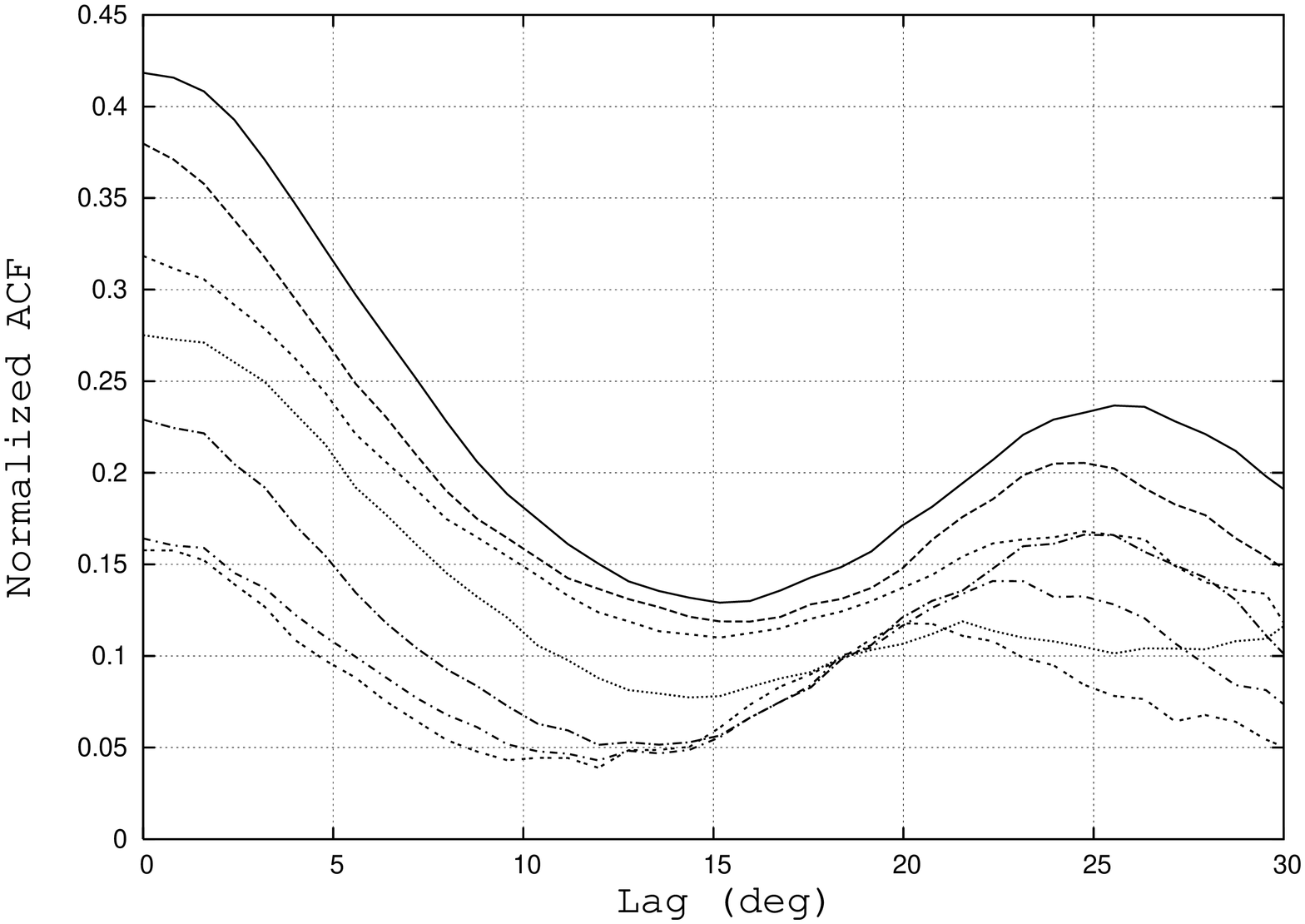}
  \includegraphics[angle=-90, width=0.45\textwidth]{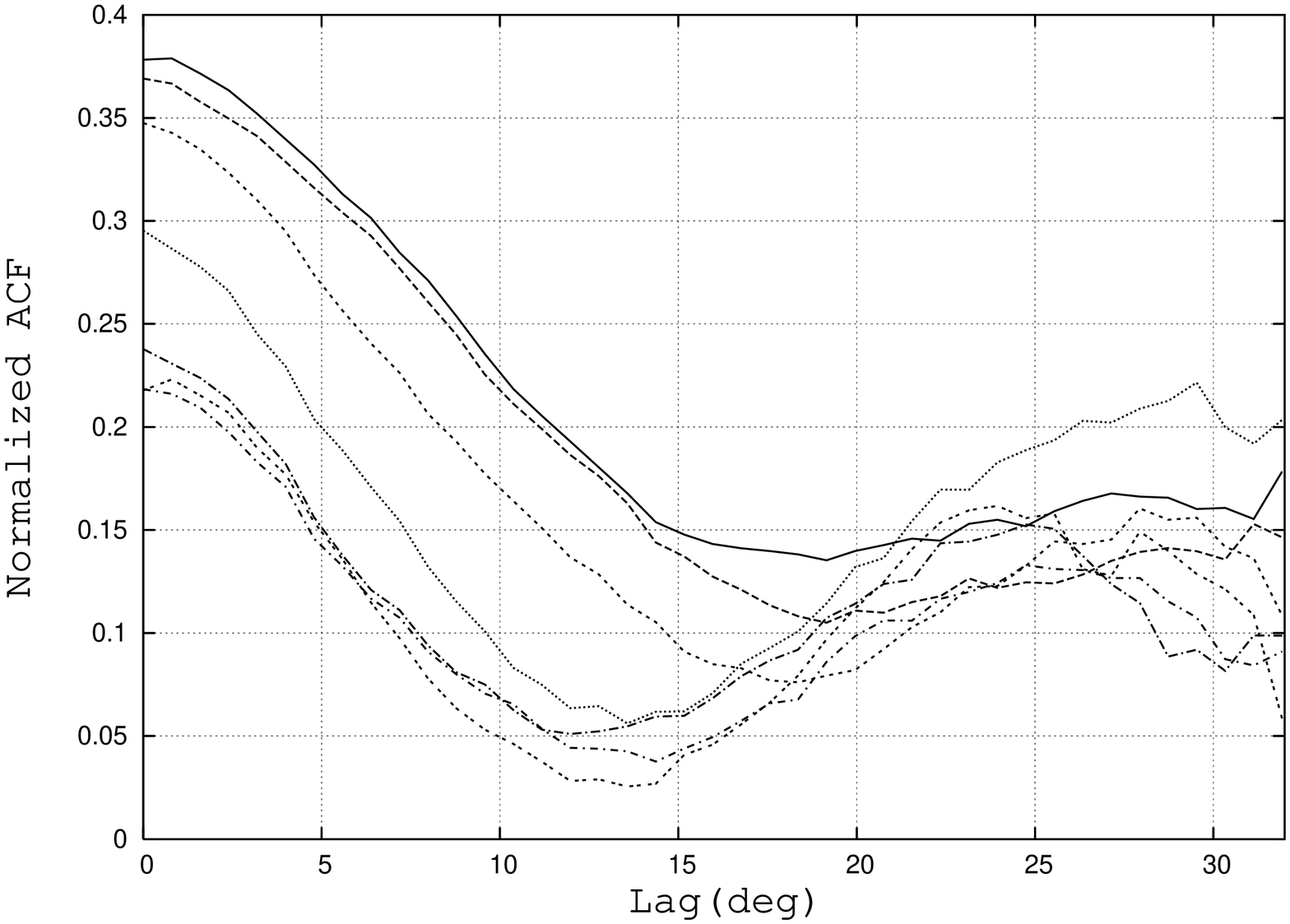}
}
\caption{Autocorrelation function for narrow pulse longitude windows at 303 MHz (left panel) and 
same at 610 MHz (right panel). Curves are for 7 different window locations: centred at 268\degr 
(solid), 260\degr (long dash), 252\degr (short dash), 244\degr (dot), 237\degr (long dash-dot), 
229\degr (short dash-dot) and 221\degr (double dot).}
\label{fig:ccf303610p2}
\end{center}
\end{figure*}
\begin{table}
\caption{ DM values of PSR B0826$-$34 from the literature survey.}
\label{table:dm_list}
\begin{tabular}{l|c|c}
\hline
Reference                                              & DM value quoted \\
                                                       & ($pc/cm^{3}$)   \\\hline
\cite{Durdin_etal}                                     & 52$^\dagger$     \\
                                                       &                 \\
EPN database                                           & 47$^\dagger$     \\
\cite{Gould_etal} \& \cite{D'Amico_etal}               &                 \\
                                                       &                 \\
\cite{Taylor_etal}                                     & 47$\pm$4$^\ddagger$ \\
                                                       &                   \\
\cite{Gupta_etal}                                      & 47$^\dagger$      \\
                                                       &                   \\
\cite{Hobbs_etal}                                      & 52.9$\pm$0.6$^\ddagger$ \\
                                                       &                 \\
\cite{Esamdin_etal}                                    & 65.6$^\dagger$    \\
                                                       &                 \\
 This paper                                            & 52.2$\pm$0.6$^\S$    \\\hline
\end{tabular}
\\
$\dagger$ : DM determination method not described \\
$\ddagger$ : DM determined using timing analysis \\
$\S$ : DM determined using  Simultaneous dual frequency observations\\
\end{table}

To study the longitude evolution of $P_2$, we repeat the correlation analysis over few 
chosen windows ($\sim$ 32\degr wide) of pulse longitude which are shifted systematically over 
a selected portion of main pulse window, in steps of 8\degr. Fig.\ref{fig:ccf303610p2} is 
a sample plot of correlation functions for 7 different pulse windows at 303 MHz and 610 MHz.
For 303 MHz we observe a smooth trend for variation of $P_2$ across the pulse window, with a 
minimum of 20.4\degr to a maximum of about 25\degr, which is compatible with the mean 
of 22.6\degr we obtained from the full main pulse.  $P_2$ increases from the centre towards the 
edge of the MP, similar to the variation reported by \cite{Gupta_etal} at 318 MHz. For 610 MHz,
the variation is from a minimum of 26.3\degr to a maximum of 30.3\degr, again with a tendency
to be lower in the middle of the MP, though the results are not as clear as for 303 MHz.
 
To study the variation of subpulse width, $\Delta\Phi_s$, with frequency, we compute the autocorrelation 
function (without any pulse offset) and find the full width at half maximum (FWHM). Average value for 
$\Delta\Phi_s$ is found to be 17.4\degr at 303 MHz, and 14.2\degr at 610 MHz, again confirming a 
frequency evolution. We collect values of $\Delta\Phi_s$ at different frequencies, from our 
observations as well as from literature, in Table. \ref{table:su_width}. We observe longitude 
evolution of $\Delta\Phi_s$: for region I (from 118\degr to 197\degr pulse longitude) 
$\Delta\Phi_s$ is significantly more than that for region II (from 197\degr to 276\degr 
pulse longitude) for all the frequencies. For a more detailed study of the longitude evolution 
of $\Delta\Phi_s$ within the MP, we followed a window scheme, similar to that employed for 
the $P_2$ variation study. We find, $\Delta\Phi_s$ varies smoothly over the pulse window, in a 
manner that is somewhat similar to that of $P_2$ variation with longitude, lower $\Delta\Phi_s$ 
values in the middle part of the MP and higher values towards the edges ($\Delta\Phi_s$ values 
near MP$_1$ being more than those near MP$_2$).   
\begin{table*}
\begin{minipage}{136mm}
\caption{ Correlation of pulse energies between different frequencies of observations.}
\label{table:cor_table}
\begin{tabular}{l|c|c|c|c}
\hline
Pulsar     & Frequency range & Correlation Coefficient & Pulse Lag up to which    & Reference              \\
           & MHz             &                         & correlation is positive &                         \\ \hline
B0031$-$07 & 275             & 70\%                    & -                       & \cite{Taylor_etal_75} \\
           & 415             &                         &                         &                         \\\hline
B0031$-$07 & 102             & 50\%                    & 10$^\dagger$            & \cite{Izvekova_etal}  \\
           & 406             &                         &                         &                         \\\hline
B0329$+$54 & 327             & 75\%                    & 2$^\dagger$             & \cite{Bartel_etal_78a} \\
           & 2695            &                         &                         &                         \\\hline
B0329$+$54 & 102             & 50$\pm$30\%             & 4$^\dagger$             & \cite{Bartel_etal_81a}  \\
           & 1720            &                         &                         &                         \\\hline
B0329$+$54$^\ddagger$ & 238  & 67$\pm$3\%              & -                       & \cite{kramer_etal}  \\
           & 626             &                         &                         &                     \\\hline
B0809$+$74 & 275             & 70\%                    & -                       & \cite{Taylor_etal_75} \\
           & 415             &                         &                         &                         \\\hline
B0809$+$74 & 102             & 30$\pm$10\%             & 4$^\dagger$             & \cite{Bartel_etal_81a} \\
           & 1720            &                         &                         &                        \\\hline
B0826$-$34 & 303             & 47\%                    & 11                      &  This paper              \\
           & 610             &                         &                         &                         \\\hline
B0834$+$06 & 275             & 70\%                    & -                       & \cite{Taylor_etal_75} \\
           & 415             &                         &                         &                         \\\hline
B1133$+$16 & 275             & 78\%                    & -                       & \cite{Taylor_etal_75} \\
           & 415             &                         &                         &                         \\\hline
B1133$+$16 & 327             & 75\%                    & 2$^\dagger$             & \cite{Bartel_etal_78a} \\
           & 2695            &                         &                         &                        \\\hline
B1133$+$16$^\S$ & 341        & 72$\pm$4\%              & -                       & \cite{kramer_etal}  \\
           & 626             &                         &                         &                     \\\hline 
B1508$+$55 & 102             & 30$\pm$20\%             & 1$^\dagger$             & \cite{Bartel_etal_81a}  \\
           & 1720            &                         &                         &                         \\\hline
\end{tabular}
- : indicate that value for the corresponding parameter is not known\\
$\dagger$ : values obtained from visual inspection of plots for crosscorrelation coefficients vs pulse lag from respective references\\
$\ddagger$ : Table 4 of \cite{kramer_etal} list the crosscorrelation coefficients between various other frequencies for B0329$+$54. \\
\cite{karastergiou_etal} reported similar values of crosscorrelation coefficients \\
$\S$ :  Table 5 of \cite{kramer_etal} list the crosscorrelation coefficients between various other frequencies for B1133$+$16\\ 
\end{minipage}
\end{table*}
\section{Discussions and Summary}                                      \label {sec:Discussions} 
\subsection {\bf Determination of accurate DM value}                   \label{sec:DM_Discussions}
Table. \ref{table:dm_list} compares our result for the DM value of PSR B0826$-$34 with other
available values from the literature $-$ these cover a wide range, with the two most recent
papers \citep{Gupta_etal,Esamdin_etal} using values of 47 and 65.6 $pc/cm^{3}$. It is clear 
that, unlike most normal pulsars, DM determination for this pulsar is a difficult and trick 
exercise, mainly because the profile is quite complex, very wide and strongly evolving with 
frequency. Given our more accurate method of determining the DM, we believe that the either 
of the extreme values above are unlikely to be correct. We note that our DM result is closest 
to value obtained by \cite{Hobbs_etal}, using multi epoch timing. The accuracy of both these 
methods depends on the choice of fiducial point. The advantage of our method of DM determination 
is that, the observations at a single epoch are self sufficient for obtaining the DM value at that 
epoch, without requiring any absolute timing information. This method can facilitate the DM 
determination of other complex and wide profile pulsars.
\subsection {\bf Study of the pulsar in null state}                    \label{null_Discussion}
We see short duration nulls of a few pulses, which appear to be simultaneous over the frequency
range of 303 to 610 MHz. In addition, this pulsar exhibits long durations of what is apparently
a null state, which also appears to be broad band over this frequency range.  
However, we find no evidence for any weak emission in these long, apparently null states of the
pulsar. Non-detection of any correlated emission structure in the null state profiles from the
simultaneous dual frequency observations (at 303 and 610 MHz) are a clear proof of this.  For 
frequencies above (up to 1060 MHz) and below (down to 157 MHz) also, we see no evidence for
any weak emission. Our best limits on this non-detection are comparable to, or a few times better, 
than the level of detection (2\%) that \cite{Esamdin_etal} report for the null state of this 
pulsar, at 1374 MHz. These comparisons are all with respect to the peak of the emission profile 
in the active state. We have also obtained absolute flux limits for the non-detection at various 
frequencies, which should be a useful comparison standard for any more sensitive studies in the 
future. 

What could be the reason for the difference between our results and those of \cite{Esamdin_etal}?
It could be a case of frequency dependent nulling, as reported by \cite{Bhat_etal} for PSR B1133$+$16 
where the pulsar exhibits cases of short duration nulls at lower frequencies like 325, 610 and 1400 MHz, 
while there is visible emission at 4850 MHz. However we believe that this explanation is difficult, 
given that we see no evidence for emission till the frequency of 1060 MHz, coupled with the fact that 
the emission properties of this pulsar appear to be significantly correlated over an octave range in 
frequency. Hence, we are unable to reconcile our results with the results of \cite{Esamdin_etal}.
\subsection {\bf Study of subpulse emission}                           \label{subpulse_Discussion}
\begin{table*}
\begin{minipage}{136mm}
\caption{\bf Frequency dependence of profile width and $P_2$ for different pulsars.}
\label{table:avp_p2}
\begin{tabular}{l|c|c|c|c|c|c}
\hline
Pulsar    & Frequency & Profile width          & $P_2$     & Comments            &  Reference  \\
          &           & $\Delta\Phi ^\diamond$ &           &                     &  \\ 
          &  MHz      &  (\degr)               & (\degr)   &                     &  \\\hline
B0031$-$07& 60        & 75           &  20.6         & $P_2\propto \nu^{-0.05}$ &\cite{Izvekova_etal}\\
          & 102       & 65           &  19.8         & $\Delta\Phi \propto \nu^{-0.4}$ &             \\
	  & 406       & 40           & 18.7         &                                 &             \\\hline
B0031$-$07& 157       & 44$\pm$5     & 18.9$\pm$1.2 & $P_2$ evolution  &\cite{Smits_etal}\\
A-mode    & 243       & 39.5$\pm$1.5 & 24$\pm$4     & does not follow    &         \\
          & 325       & 39.4$\pm$0.3 & 19.8$\pm$1.0 & definite trend  &         \\
          & 607       & 32$\pm$4     & 21$\pm$3     &                 &         \\
          & 840       & 32.2$\pm$0.8 & 10$\pm$3     &                 &         \\
          & 1167      & 25$\pm$6     & 14$\pm$5     &                 &         \\
          & 4850      & 23.6$\pm$0.6 & 14$\pm$2     &                 &         \\\hline
B0031$-$07& 157       & 46.3$\pm$1.4 & 19.2$\pm$0.7 & $P_2$ decreases &\cite{Smits_etal}\\
B-mode    & 243       & 42.3$\pm$0.6 & 18.7$\pm$1.6 & or remains unchanged  &         \\
          & 325       & 40.3$\pm$0.12& 18.7$\pm$0.6 & with increasing      &         \\
          & 607       & 39.3$\pm$0.9 & 18.7$\pm$0.6 & frequency            &         \\
          & 840       & 43$\pm$0.9   & 14$\pm$1.6   &                      &         \\
          & 1167      & 38$\pm$5     & 12$\pm$3     &                      &         \\\hline
B0320$+$39& 102       & 11           & 3.8          & $P_2 \propto \nu^{-0.15}$      &\cite{Izvekova_etal}\\
          & 406       & 8.5          & 3.1          & $\Delta\Phi \propto \nu^{-0.4}$&             \\\hline
B0809$+$74& 102       & 34.8         & 14.7         & see Sect.\ref{subpulse_Discussion} &\cite{Davies_etal} \\
          & 406       & 26.0         & 10.8         &                 & \\
          & 1412      & 29.8         & 8.6          &                 & \\
          & 1720      & 27.2$^b$     & 8.1          &                 & \cite{Bartel_etal_81a}\\\hline
B0826$-$34& 157       & 140$\pm$4$^\dagger$  &  23 $\pm$0.8  &                                   &   This Paper        \\
          & 303       & 132$\pm$4$^\dagger$  &  22.6$\pm$0.8 & $\Delta\Phi \propto \nu^{-0.2}$   & This paper \\
          & 318       & 134$\pm$4$^\dagger$  &  24.9$\pm$0.8 & $P_2$ increases with               & \cite{Gupta_etal} \\
          & 610       & 116$\pm$4$^\dagger$  &  27.9$\pm$0.8 & increasing frequency from         & This paper \\
          & 645       & 112$\pm$4$^\dagger$  &  29$\pm$2.0    &303 to 645 MHz                   &\cite{Biggs_etal}\\
          & 1060      & 117$\pm$4$^\dagger$  &  28.7$\pm$0.8 &                                 &  This Paper         \\
          & 1374      &  -                   &  27.5         &                                 &\cite{Esamdin_etal} \\\hline
B1237$+$25& 430       & 16$^a$       & 10.6$\pm$2   & $P_2 \propto \nu^{-0.16}$  &\cite{Wolszczan_etal}\\
          & 1700      & 14.5$^b$     & 9.1$\pm$1.5  & $\Delta\Phi \propto \nu^{-0.07}$&          \\\hline
B2016$+$28& 102       & 20           & 8.5          & $P_2 \propto \nu^{-0.1}$       &\cite{Izvekova_etal}\\
          & 406       & 15           & 6.2          & $\Delta\Phi \propto \nu^{-0.2}$&                    \\
          & 1412      & 15           & 5.5          &                                &               \\\hline
B2020$+$28& 430       & 17$^a$       & 28.3$\pm$4   & $P_2$ increases with   &\cite{Wolszczan_etal} \\
          & 1700      & 16.2$^b$     & 51.4$\pm$7   & frequency             &                     \\\hline
\end{tabular}
\\
$\diamond$ : Profile width at 10\% intensity level\\
a : average profile width at 400 MHz from \cite{Gould_etal}\\
b : average profile width at 1600 MHz from \cite{Gould_etal}\\
I,II : left and right part of MP respectively, see Fig.1 of \cite{Biggs_etal}\\
$\dagger$: separation between two components of main pulse\\
\end{minipage}
\end{table*} 
(i) Pulse energy correlations : Our result that the fluctuations of the main pulse energies at
303 and 610 MHz are correlated up to 47\% level, indicates that the emission process generating
the single pulses has a high degree of correlation over a this octave bandwidth. What is equally
noteworthy is that this level of correlation is pretty much the same as that at each of these
individual frequencies, indicating that there is practically no loss of correlation over this
range of frequency. Table.\ref{table:cor_table} lists the correlation coefficient of pulse
energies for different pulsars between two different observing frequencies, reported in the
literature. All the pulsars listed there show significantly high correlation, supporting the
case for the broad band nature of pulsar radio emission. At the same time, the significantly
lower correlation between the main pulse and inter-pulse energies at 610 and 1060 MHz, implies
that the physical processes responsible for the  MP and IP emission may be much more unrelated
to each other.

The result that the intensity correlations are positive for large lags implies that there is
some kind of memory in the underlying subpulse structure. That this memory is the longest for
PSR B0826$-$34 (amongst all known cases), is quantified in column 4 of Table.\ref{table:cor_table}.
This is likely to be related to the large number of subpulses (or drift bands) that each main
pulse energy estimate encompasses. The memory could then correspond to the time taken for the
emission from a given subpulse spark column to drift through the main pulse window.

The other memory that the main pulse intensity variations exhibit is the intriguing feature
of the peak at lag 13 seen in some of the correlation results. This could be an indication of
the total time for rotation, around the magnetic axis, of the pattern of drifting subpulses $-$
the carousel rotation time. For example, in the 14 spark model presented by \cite{Gupta_etal},
the drift rate is such that one subpulse drifts to a location close to that of the adjacent
subpulse in one pulsar rotation, leading to a carousel rotation time of $P_4=14 P_1$.
However, if we believe that there are 13 sparks, as indicated in the model given by
\cite{Esamdin_etal}, with the same drift rate condition as proposed by \cite{Gupta_etal}, then
we would have $P_4=13 P_1$, matching with the secondary peak feature seen in our data.

(ii) Study of subpulse drifting:
PSR B0826$-$34 exhibits substantial variations of $P_2$ at different frequencies, usually reported
at different observing epochs (Table.\ref{table:avp_p2}). The different values of $P_2$ determined
from the simultaneous data at 303 and 610 MHz strongly supports the claim that these are genuine
variations with frequency, and not due to some time dependent phenomenon. The collected results
in Table.\ref{table:avp_p2} are consistent with a scenario where $P_2$ increases with frequency
from 157 to 645 MHz, but then appears to saturate or even turn over at the higher frequencies like
1060 and 1374 MHz. For the low frequency range, we find that this frequency evolution
($P_2 \propto \nu^{0.3}$) is of opposite sense to that for the profile width, as measured by the
separation between two components of the MP ($\Delta\Phi \propto \nu^{-0.2}$). Only two other
pulsars in Table.\ref{table:avp_p2} show such an abnormal behaviour : PSR B2020$+$28 shows
a clear opposite sense (though only 2 frequency points are available) and PSR B0031$-$07 in
the A-mode shows erratic increases and decreases of $P_2$ with increasing frequency.

For most pulsars, though both $P_2$ and the average profile width decrease with increasing
frequency, the evolution of these two quantities with frequency does not exactly match (see
Table.\ref{table:avp_p2}). \cite{Izvekova_etal} showed in detail that $P_2$ is, in general,
less dependent on frequency, as compared to the profile width $-$ this is reflected in our
Table.\ref{table:avp_p2} as well. This has been explained \citep{Gil_etal_96,Gil_etal} to be a 
natural consequence of the conal model of pulsar beams \citep{Rankin_etal_93} $:$ the subpulse 
enhancement follows a narrow bundle of dipole field lines, while the mean enhancement corresponding 
to the peaks of mean profile components, is distributed over a cone (or a number of nested cones) 
of dipole field lines. Thus, the frequency dependence of subpulses and pulse widths should in 
general be different. However, these models do not appear to explain the opposite sense that we see in PSR B0826$-$34.

Another pulsar with anomalous frequency variations of profile width and $P_2$ is PSR B0809$+$74.
$P_2$ being 1:8 times greater at 102.5 MHz than at 1720 MHz, for this pulsar, is explained 
by \cite{Edwards_etal} as an artifact of the phase step, which is only present at high frequencies. 
Whereas \cite{Rankin_etal_05} analysed the $P_2$ variations between 40 to 1400 MHz 
for this pulsar, and showed that they arise from incoherent superposition of two orthogonal modes 
of polarisation. It will be interesting to see if detailed polarisation studies of PSR B0826$-$34 
can throw some light on the $P_2$ variations with frequency.

The other interesting result is the variation of $P_2$ with pulse longitude, where we confirm
the behaviour first reported by \cite{Gupta_etal} at 318 MHz (a variation from 21.5\degr to
27\degr within the MP), and also find some signs for a similar effect at 610 MHz.  Similarly,
\cite{Esamdin_etal} reported that at 1374 MHz, $P_2$ varies from 26.8\degr to 28\degr within
the MP. \cite{Gupta_etal} have provided an explanation of this in terms of the geometrical
effect of the way our line-of-sight traverses the polar cap.

$\Delta\Phi_s$ determined in this work is compared to the result quoted by \citep{Biggs_etal,Esamdin_etal} 
at 645 and 1374 MHz respectively in Table. \ref{table:su_width}. $\Delta\Phi_s$ is observed to be frequency 
dependent, average $\Delta\Phi_s$ at 303 MHz is more than 610 MHz value (see Sect.\ref{subpulse_study}). In 
addition, $\Delta\Phi_s$ is observed to be longitude dependent: $\Delta\Phi_s$ in region I is more than region II 
(Table. \ref{table:su_width}). Similar longitude evolution of $\Delta\Phi_s$ within the MP is reported by 
\cite{Biggs_etal} and \cite{Esamdin_etal}.

To summarise, using results from high sensitivity single pulse observations simultaneously at
303 and 610 MHz, and individually at 157, 325, 610 and 1060 MHz, we have investigated different
aspects of PSR B0826$-$34: (1) precise value of DM  (2) emission properties of the long duration
null states and (3) simultaneous subpulses and details of the drifting behaviour.
\begin{table}
\caption{Frequency dependence of subpulse width ($\Delta\Phi_s$) of PSR B0826$-$34.}
\label{table:su_width}
\begin{tabular}{l|c|c|c|c|c|c|c}
\hline
Frequency & \multicolumn{4}{|c|}{$\Delta\Phi_s$} & Reference     \\
  (MHz)   & \multicolumn{4}{|c|}{(\degr)}   &                \\
              &  MP          & I         &  II       &   IP           &                \\ \hline
157$^\ddagger$&  15.8$^\S$   & 22.1$^\S$ & 10.2$^\S$ &  -             & This paper               \\
303$^\dagger$ &  17.4$^\S$   & 28.4$^\S$ & 12.6$^\S$ &  -             & This paper               \\
610$^\dagger$ &  14.2$^\S$   & 22$^\S$   & 12.6$^\S$ & 12.6$^\S$      & This paper               \\
610$^\ddagger$&  14.2$^\S$   & 22.9$^\S$ & 12.6$^\S$ & 12.6$^\S$      & This paper               \\
645           &  -           & 14$^\ast$ & 5.9$^\ast$& 4  $^\ast$     & \cite{Biggs_etal}        \\
1060$^\ddagger$& 14.2$^\S$   & 20.5$^\S$ & 12.6$^\S$ & 17.4$^\S$      & This paper               \\
1374           & 13.1$\pm$1.9&    -      &    -      & 19$\pm$3.5     & \cite{Esamdin_etal}      \\\hline
\end{tabular}
\\
$\dagger$ : From simultaneous dual frequency observing session \\
$\ddagger$ : From single frequency observing session \\
I : 118\degr to 197\degr  pulse longitude\\
II : 197\degr to 276\degr  pulse longitude\\
$\ast$ : Lag corresponding to half width at half maximum (HWHM) values\\
$\S$ : Error associated in determination of $\Delta\Phi_s$ is $\pm$1.6\degr\\
\end{table}

\section{Acknowledgments}
We thank the stuff of the GMRT for help with the observations. The GMRT is run by the National
Centre for Radio Astrophysics of the Tata Institute of Fundamental Research.  JG acknowledges the
partial support of Polish Grant N N203 2738 33.


\label{lastpage}       
\end{document}